\documentclass[aps,prd,twocolumn,showpacs,amsmath,amssymb]{revtex4-1}
\usepackage{epsfig}
\usepackage{graphicx}
\usepackage{dcolumn}
\usepackage{bm}
\usepackage{overpic}
\usepackage{subfigure}
\usepackage{float}
\usepackage{color}
\usepackage{amsmath}
\usepackage{mathcomp}
\usepackage{multirow}
\usepackage[dvipdfm, colorlinks,linkcolor=blue, citecolor=blue, urlcolor=black]{hyperref}
\usepackage{rotating}

\newcommand{\EE}{e^+e^-}

\newcommand{\jpsi}{J/\psi}

\def\Journal#1#2#3#4{{#1} {\bf #2}, #3 (#4)}

\def\PLB{Phys. Lett. B}
\def\PRL{Phys. Rev. Lett.}
\def\PRD{Phys. Rev. D}

\def\EPJA{Eur. Phys. J. A}

\begin{document}
\normalsize
\parskip=5pt plus 1pt minus 1pt

\title{\boldmath An overview of $\eta_c(1S)$, $\eta_c(2S)$ and $h_c(1P)$ physics at BESIII}

\author{Qingping Ji$^{a}$}
\email[E-mail: ]{jiqingping@htu.edu.cn}
\affiliation{Henan Normal University, Xinxiang 453007, China}
\author{Shuangshi Fang$^{b}$}
\email[E-mail: ]{fangss@ihep.ac.cn}
\affiliation{Institute of High Energy Physics, Chinese Academy of Science, Beijing 100049, China}
\author{Zhiyong Wang$^{b}$}
\email[E-mail: ]{wangzy@ihep.ac.cn}
\affiliation{Institute of High Energy Physics, Chinese Academy of Science, Beijing 100049, China}

\vspace{4cm}

\date{\today}

\begin{abstract}

With the help of the largest data samples of $J/\psi$ and $\psi(2S)$ events ever produced in $e^+e^-$ annihilations, the three singlet charmonium states, $\eta_c(1S)$, $\eta_c(2S)$ and $h_c(1P)$, have been extensively studied at the BESIII experiment. In this review, a survey on the most recent results, including a series of precision measurements and observations of their new decay modes, is presented, which indicates the further investigations on their decays are needed to understand their decay mechanisms and have precision tests of the theoretical models.  At present, about eight times larger data samples of 10 billion $J/\psi$ events and 3 billion $\psi(3686)$ events were collected with the BESIII detector, and thus the prospects for the study of these three charmonium states is discussed extensively.

\end{abstract}

\pacs{13.20.Pq, 13.25.Gv, 12.38.Qk}

\maketitle

\clearpage
\clearpage

\section{Introduction}
\label{sec:introduction}

\begin{figure}[hbtp]
\centering
\includegraphics[height=7cm,width=7.9cm]{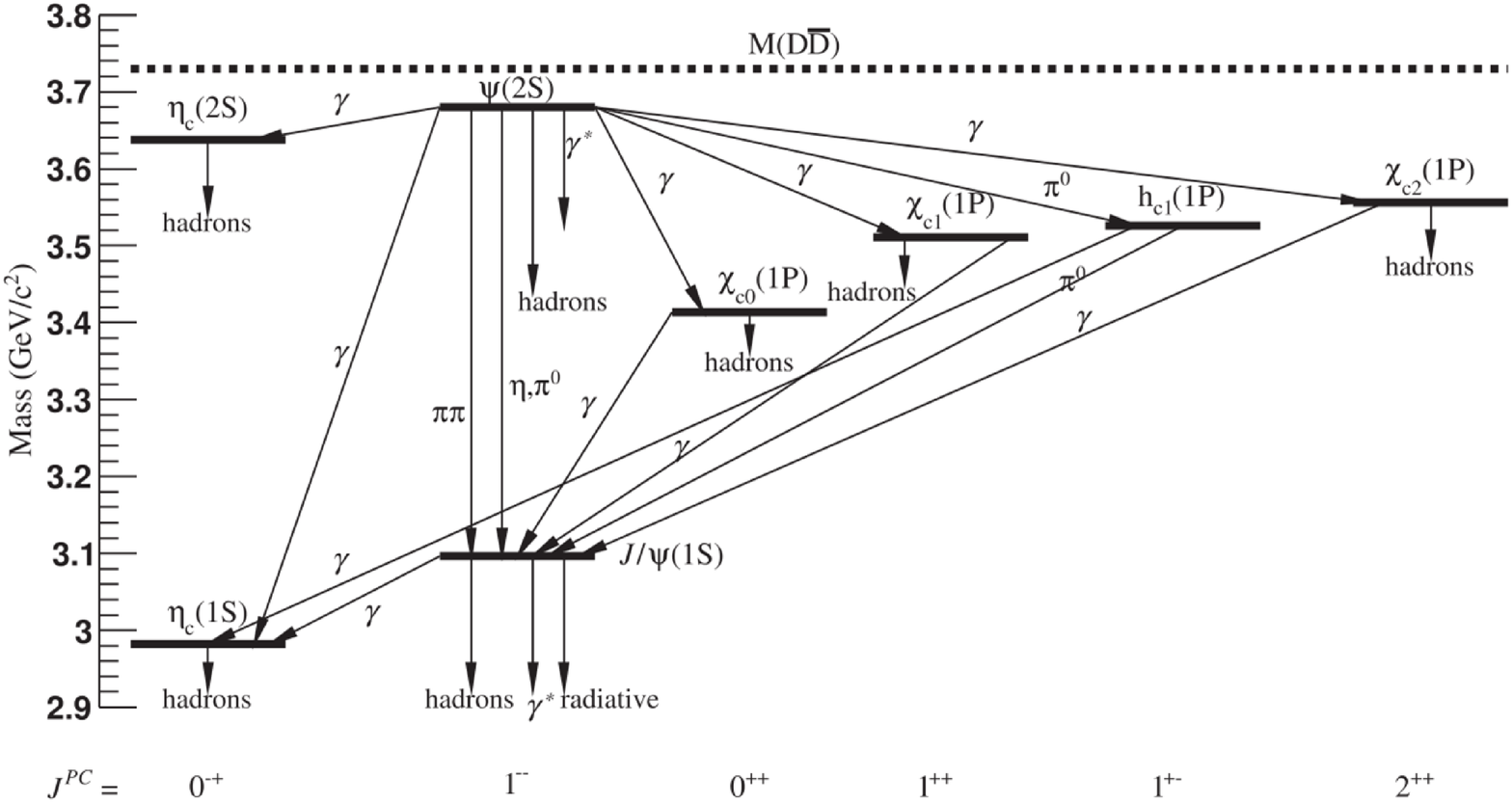}
\caption{Low-lying charmonium ($c\overline{c}$) spectrun, with selected decay modes and transition~\cite{newpsip}.
}
\label{charmonium}
\end{figure}

About four decades ago, the discovery of  the $J/\psi(1S)$ not only revealed the existence of charm quark, but led to the advent of the charmonium spectroscopy
indicated by Fig.~\ref{charmonium}, which  is of particular interest since it provides a bridge for studying the dynamics of quantum chromodynamics(QCD) in the interplay of perturbative and non-perturbative QCD regime.

Soon after that the lowest-lying $S$-wave spin-singlet charmonium state, $\eta_c(1S)$, was  discovered in the radiative transitions of  $\psi(2S)$ and $J\psi$~\cite{etac1S}. Naturally the experimental searches for  its  radial excited state, $\eta_c(2S)$,  were performed accordingly, but no evident signal was observed in $\psi(2S)$ radiative decays \cite{searchEtac2SCB,searchEtac2SBES1,searchEtac2SBES2,searchEtac2SCLEO}. Until 2002, the $\eta_c(2S)$ was discovered in $B$ meson decays~\cite{etac2S} by the Belle experiment, and subsequently confirmed in the processes of two-photon and the double-charmonium production processes~\cite{etac2S1,etac2S2, etac2S12,etac2S12}.  The measured
mass,$3637.7\pm1.3$ MeV/$c^2$~\cite{pdg}, is quite close to the mass of $\psi(2S)$, which indicates that the detector should have a very excellent resolution for the low energy photon
in $\psi(2S)$ radiative transition to investigate $\eta_c(2S)$ decays.

For the $P$-wave singlet state, $h_c(1P)$,  it also have been searched for several decades. In 1992, E760 reported the first evidence  in the process of $p\bar{p}\to h_c(1P)\to\pi^0 J/\psi$~\cite{hc}, but subsequently excluded by the E835 experiment with a larger data sample~\cite{hc1}.
After that E835 claimed the observation of the $h_c(1P)$ in the process of $p\bar{p}\to h_c(1P)\to\gamma\eta_c(1S)$~\cite{hc1}. Soon it was confirmed by the CLEO collaboration ~\cite{hc2,hc3} in the decays of $\psi(2S)\to\pi^0 h_c(1P)$ with $h_c(1P)\to\gamma\eta_c(1S)$.  And an evidence for $h_c(1P)$ decays to multi-pion final states~\cite{hc4} was also reported.

From the  discoveries of $\eta_c(1S)$, $\eta_c(2S)$ and $h_c(1P)$,  to have a precision study of their properties in $J/\psi$ or $\psi(2S)$ decays, both the large data samples and the
excellent performance of the detector are   strongly needed.  And the BESIII experiment~\cite{bes3}
 fulfills these requirements and provides a unique opportunity to investigate their decays.
With the large data samples of $1.31\times10^9$ $J/\psi$ events~\cite{NJpsi1, NJpsi2} and $448.1\times10^6$ $\psi(2S)$ events~\cite{Npsip1,Npsip2} are collected with the BESIII detector,
The available $\eta_c(1S)$ and $h_c(1P)$ events via transition decays of $J/\psi(1S)\to\gamma\eta_c(1S)$, $\psi(2S)\to\gamma\eta_c(1S)$ and $\psi(2S)\to\pi^0 h_c(1P)$ are summarized in Table.~\ref{tab::hcsample}. And a series of achievements on  the study of $\eta_c(1S)$, $\eta_c(2S)$ and $h_c(1P)$ was achieved, including precision measurements and observations of new decays.


In this article, we first have a brief review on the theoretical underpinning in Sec.~\ref{sec:theoretical} discussing potential models and radiative transitions, then
review the progress on studying $\eta_c(1S)$, $\eta_c(2S)$ and $h_c(1P)$ decays at BESIII experiment in Sec.~\ref{sec:etacdecays}, Sec.~\ref{sec:etac2Sdecays} and Sec.~\ref{sec:hcdecays}, respectively. 
Finally, we discuss the prospect in the future studies on these charmonium states at BESIII experiment in Sec.~\ref{sec:summary}.

\begin{table*}[hbtp]
\begin{center}
\caption{The available $\eta_c(1S)$, $\eta_c(2S)$ and $h_c(1P)$ decays calculated with the $1.31\times10^9$ $J/\psi$ events, $448.1\times10^6$ $\psi(2S)$ events and {\sc XYZ}~\cite{xyzparticle} data sample at BESIII. Here, we only list the $h_c$ events at $\sqrt{S}$ = 4.23 GeV.}
\begin{tabular}{lcc}
\hline
Decay mode                           & ${\cal{B}}$~\cite{pdg}/$\sigma$~\cite{Zc40420}      & $\eta_c(1S)$, $\eta_c(2S)$/$h_c(1P)$ events \\
\hline
$J/\psi(1S)\to\gamma\eta_c(1S)$          &  $(1.7\pm0.4)\times10^{-2}$             & $2.2\times10^7$\\
$\psi(2S)\to\gamma\eta_c(1S)$      &  $(3.4\pm0.5)\times10^{-3}$  & $1.5\times10^6$\\
$\psi(2S)\to\gamma\eta_c(2S)$      &  $(7.0\pm5.0)\times10^{-4}$      & $3.1\times10^5$\\
$\psi(2S)\to\pi^0 h_c(1P)$         &  $(8.6\pm1.3)\times10^{-4}$  & $3.9\times10^5$\\
$e^+e^-\to\pi^+\pi^- h_c(1P)$         &  $(50.2\pm2.7\pm4.6\pm7.9)\rm pb$  & $2.3\times10^4$\\
\hline
\end{tabular}
\label{tab::hcsample}
\end{center}
\end{table*}

\section{Theoretical underpinning}
\label{sec:theoretical}

The lowest-lying and well-established charmonium states below the open-charm threshold can be used to precisely test predictions based on Quantum chromodynamics (QCD) and QCD-inspired models in a region where both perturbative (PQCD) and non-perturbative (NPQCD) aspect paly a role. 
Determining the internal structure to previous established charmonium states, measuring masses and widths with high precision, precisely measuring transitions (both radiative and hadronic)
between charmonium states and finding new decay modes can provide a unique and important perspective on the dynamics of strong force physics.

\subsection{Potential model}\label{sec:potential}
\subsubsection{Nonrelativistic potential model}
Nonrelativistic (NR) potential model is a minimal model of the charmonium system, based on the wave functions determined by the Schr$\ddot{\rm o}$dinger equation with a conventional quarkonium potential.
In NR potential model, the central potential is
\begin{widetext}
\begin{eqnarray}
V_0(r)=-\frac{4}{3}\frac{\alpha_s}{r}+br+\frac{32\pi\alpha_S}{9m_c^2}\tilde{\delta}_{\sigma}(r)\vec{S}_c\cdot \vec{S}_{\overline{c}}
 + \frac{1}{m^2_c}[(\frac{2\alpha_s}{r^3}-\frac{b}{2r})\vec{L}\cdot\vec{S}+\frac{4\alpha_s}{r^3}T],\label{eq1}
\end{eqnarray}
\end{widetext}
in which the spin-spin operator $(\vec{S}\cdot\vec{S})$ is for the mass splitting between spin-triplet and spin-singlet state, spin-orbit operator $(\vec{S}\cdot\vec{L})$ and tensor operator $T$ are for
the mass splitting between spin-triplet states~\cite{potenModel}.  Measurement the mass of charmonium state with high precision would be helpful to understanding the contribution form the spin associated.

\subsubsection{Godfrey-Isgur relativized potential model}
The Godfrey-Isgur (GI) model, which is a ``relativized'' extension of the nonrelativistic model, assumes a relativistic dispersion relation for the quark kinetic energy, a QCD-motivated running coupling
$\alpha_s(r)$, a flavor-dependent potential smearing parameter $\sigma$, and replaces factors of quark mass with quark kinetic energy. Details of the model and the method of solution may be found
in Ref.~\cite{GI}.

According to the Refs.~\cite{potenModel}~\cite{GI}, one important aspect about the two models is that the GI model gives reasonable accurate results for the spectrum and matrix elements of quarkonia of all $u$, $d$, $s$, $c$, $b$ quark flavors, whereas
the NR model is only fitted to the $c\overline{c}$ system.

\subsection{Radiative Transitions}

Transition ratio between charmonium states are calculated based on different potentials (nonrelativistic model or relativistic effect correction considered). Mass difference between charmonium spin-triplet and spin-singlet state, as well as the transition ratio are also calculated with lattice QCD (LQCD)~\cite{Lqcd1}~\cite{Lqcd2}. Measurement the transition ratios could test the these predictions.

\subsubsection{Electric dipole Transitions}
The partial widths for electric dipole ($E1$) transitions is evaluated as~\cite{potenModel}

\begin{widetext}
\begin{eqnarray}
\Gamma_{E1}(n^{2S+1}L_{J}\to n^{\prime2S\prime+1}L^{\prime}_{J^{\prime}}+\gamma)=\frac{4}{3}C_{fi}\delta_{SS^{\prime}}e^2_c\frac{\alpha}{m^2_c}
\arrowvert \langle\psi_f|r|\psi_i\rangle\arrowvert^2 \times E^3_\gamma\frac{E^{(c\overline{c})}_f}{M^{(c\overline{c})}_i},\label{eq1-1}
\end{eqnarray}
\end{widetext}
where $e_c=2/3$ is the $c$-quark charge in units of $\arrowvert e\arrowvert$, $\alpha$ is the fine-structure constant, $E_{\gamma}$ is the fine photon energy, $E^{(c\overline{c})}_{f}$
is the total energy of the final $c\overline{c}$ state, $M^{(c\overline{c})}_i$ is the mass of the initial $c\overline{c}$ state, the spatial matrix element
$\arrowvert \langle\psi_f|\psi_i\rangle\arrowvert$ involves the initial and final radial wave functions, and the angular matrix element $C_{fi}$ is
\begin{eqnarray}
C_{fi}={\rm max}(L,L^{\prime})(2J^{\prime}+1)C^2_m,
\label{eq1-3}
\end{eqnarray}
with
\begin{gather*}
C_m=
\begin{Bmatrix} L^{\prime} & J^{\prime} & S \\ J & L & 1 \end{Bmatrix}\\
\end{gather*}\label{eq1-4}
(See the previous $E1$ formula for definitions.)

Transitions from initial $1^{--}$ $c\overline{c}$ states are of greatest interest since these can be studied with high statistics at $e^+e^-$ machines, such as $\psi(2S)\to\gamma\chi_{c0,1,2}(1P)$, and the transitions from the $\psi(4040)$ and $\psi(4415)$

\subsubsection{Magnetic dipole Transitions}
Although Magnetic dipole transition ($M1$) rates are typically rather weaker than $E1$ rates, they are nonetheless interesting because they may allow access to spin-singlet states that are very difficult to
produce otherwise. It is also interesting that the known $M1$ rates show serious disagreement between theory and experiment. This is in part due to the fact that $M1$ transitions
between different spatial multiplets, such as $\psi(2S)\to\gamma\eta_c(1S)$, $J/\psi(1S)\to\gamma\eta_c(1S)$ and $\psi(2S)\to\gamma\eta_c(2S)$, are nonzero only due to small relative corrections to a vanishing
lowest-order $M1$ matrix element.

The $M1$ radiative partial widths are evaluated using~\cite{potenModel}
\begin{widetext}
\begin{eqnarray}
\Gamma_{M1}(n^{2S+1}L_{J}\to n^{\prime2S\prime+1}L^{\prime}_{J^{\prime}}+\gamma)=\frac{4}{3}\frac{2J^{\prime}+1}{2L+1}\delta_{LL^{\prime}}\delta_{S,S^{\prime}\pm1}e^2_c\frac{\alpha}{m^2_c}
\arrowvert \langle\psi_f|\psi_i\rangle\arrowvert^2 \times E^3_\gamma\frac{E^{(c\overline{c})}_f}{M^{(c\overline{c})}_i},\label{eq1-2}
\end{eqnarray}
\end{widetext}

In Ref.~\cite{potenModel}, the $E1$ rates, as well as the $M1$ rates are evaluated in both the NR potential model and the GI model described in Ref.~\cite{GI}.
Better experimental data will be very important for improving our description of these apparently simple but evidently poorly understood $M1$ radiative transitions.


\section{$\eta_c(1S)$ physics}
\label{sec:etacdecays}
\subsection{$\eta_c(1S)$ resonant parameters}
Although $\eta_c(1S)$ has been known for about thirty years~\cite{etac1S}, its resonant parameters are still have large uncertainties when compared to those of other charmonium states~\cite{pdg}.
Early measurements about the $\eta_c(1S)$ properties using $J/\psi(1S)$ radiative transition~\cite{etac1SMark3,etac1SBES} found its mass and width to be about 2980 MeV/$c^2$ and 10 MeV, respectively. However, the experiments including photon-photon fusion and $B$ decays have appeared a significantly
discrepancies on both mass and width~\cite{etac1Scleo,etac1Sbabar,etac1Sbelle,etac1Sbelle2}. In 2009, CLEO Collaboration pointed out a distortion of the $\eta_c(1S)$ line shape in $\psi$ decays using the $M1$ transition both $J/\psi(1S)\to\gamma\eta_c(1S)$
and $\psi(2S)\to\gamma\eta_c(1S)$~\cite{etac1Scleo2}. They attributed this distortion to the energy dependence of the $M1$ transition matrix element. The $h_c(1P)\to\gamma\eta_c(1S)$ transition can provide a new laboratory to study $\eta_c(1S)$ properties since the $\eta_c(1S)$ line shape in this transition should be generally normal due to the small non-resonant interfering backgrounds.

Better measurements of its mass and total width are still inspired.
These measurements are performed via $M1$ transition ($\psi(2S)\to\gamma\eta_c(1S)$~\cite{etac1SBES3} and $J/\psi(1S)\to\gamma\eta_c(1S)$~\cite{etac1S2ww}) and $E1$ transition decay, $h_c(1P)\to\gamma\eta_c(1P)$ via $\psi(2S)\to\pi^0 h_c(1P)$~\cite{etac1S216} involving $\eta_c(1S)$ at BESIII.

\begin{figure*}[hbtp]
\centering
\epsfig{width=0.9\textwidth,clip=true,file=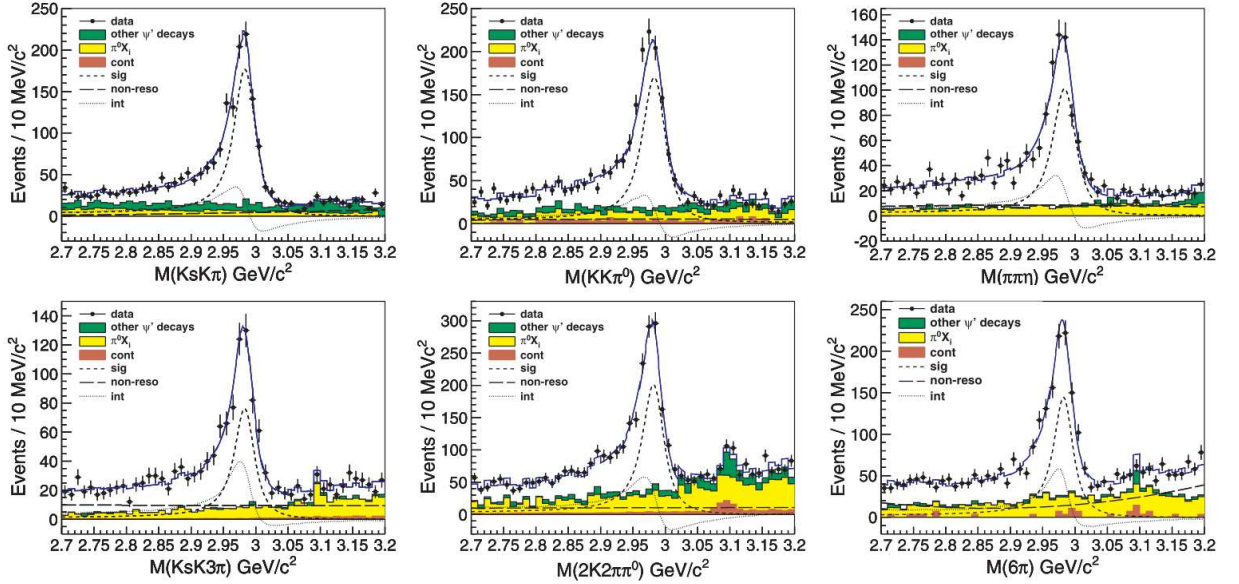}\\
\caption{Measurement of the mass and width of $\eta_c(1S)$ using $\psi(2S)\to\gamma\eta_c(1S)$.  The invariant mass distributions for the decays $K_S K^+\pi^-$, $K^+ K^-\pi^0$, $\eta\pi^+\pi^-$, $K_S K^+\pi^+\pi^-\pi^-$, $K^+ K^-\pi^+\pi^-\pi^0$, and $3(\pi^+\pi^-)$, respectively, with the fit results (for the constructive solution) superimposed~\cite{etac1SBES3}. Dots with error bars are data, and the curves are total
fit and each component. Charge conjugate modes are included.}
\label{etac1S}
\end{figure*}

\begin{figure*}[hbtp]
\centering
\includegraphics[height=4cm,width=5.3cm]{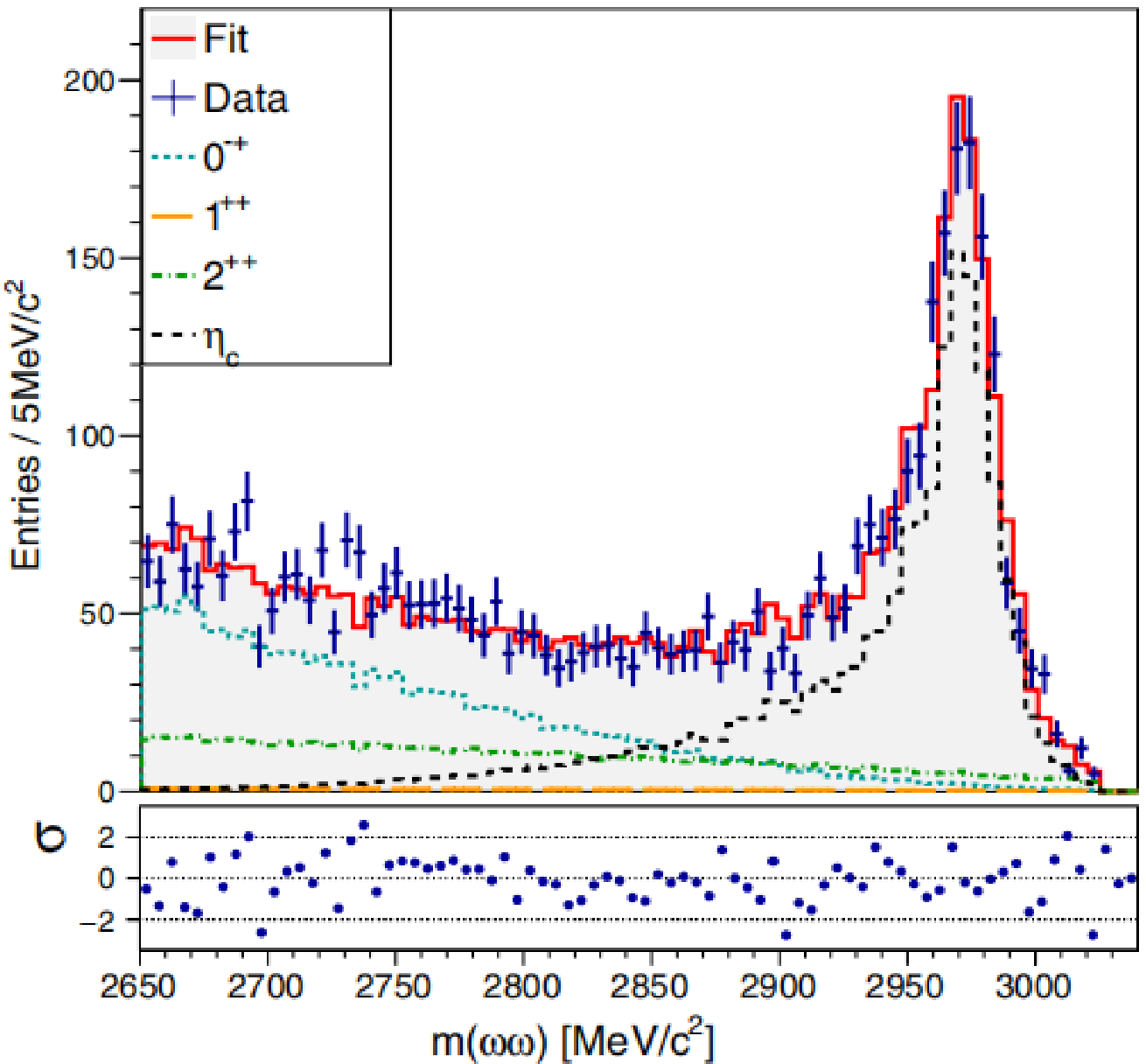}
\put(-90,100){\bf(a)}
\includegraphics[height=4cm,width=5.3cm]{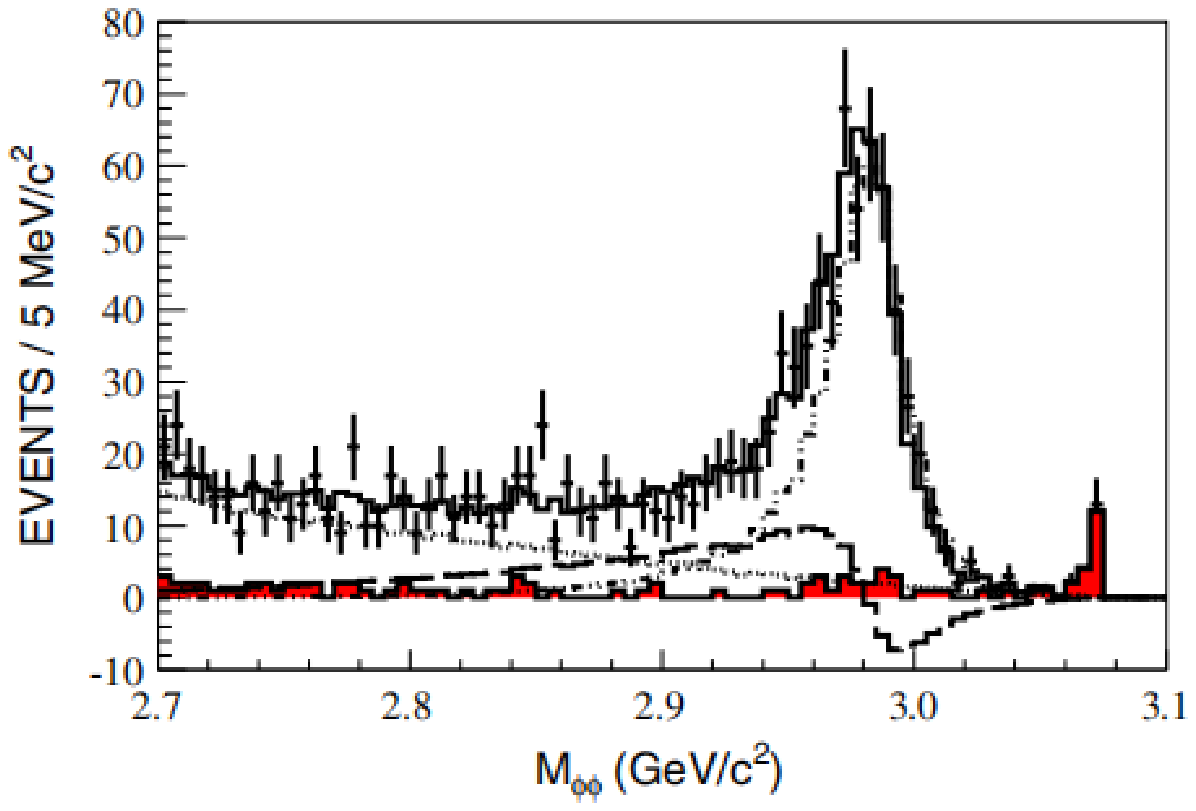}
\put(-90,100){\bf(b)}
\includegraphics[height=4cm,width=5.3cm]{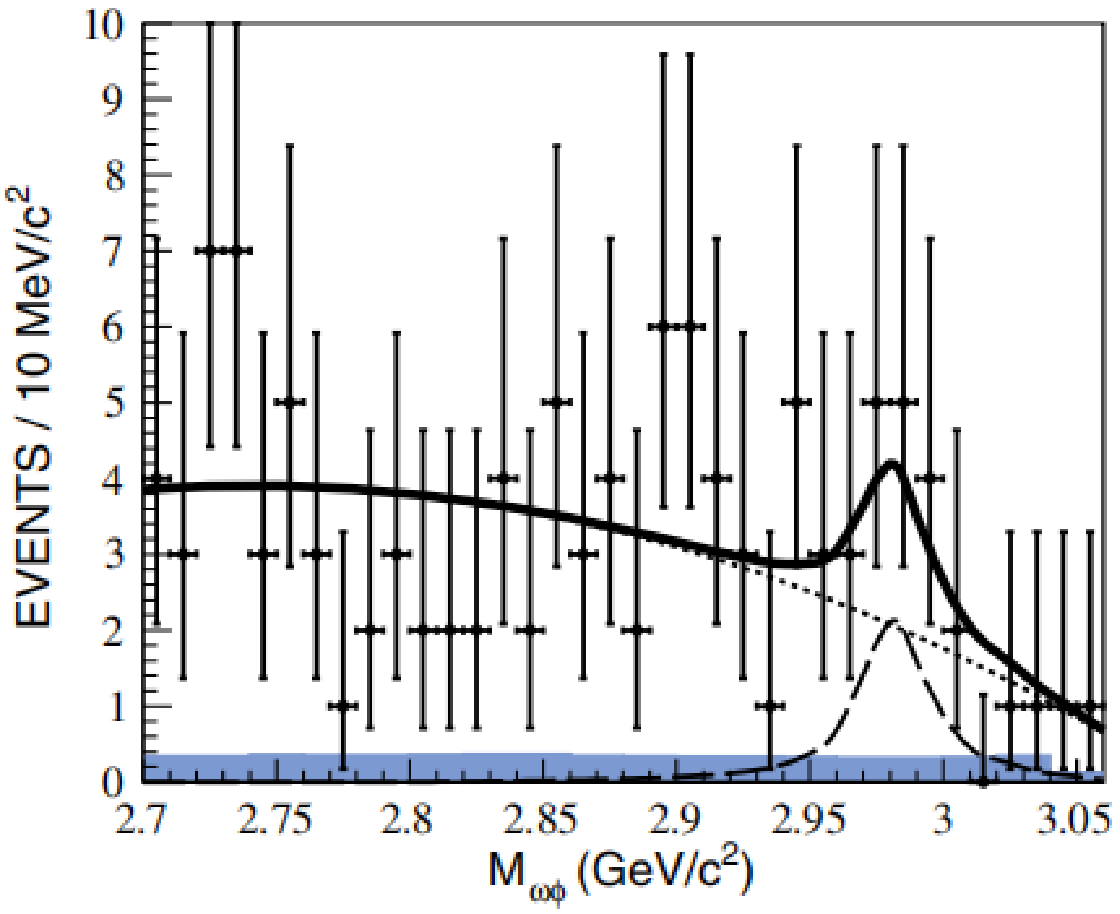}
\put(-90,100){\bf(c)}
\caption{Observation of $\eta_c(1S)\to\omega\omega$~\cite{etac1S2ww}, and improved measurements of BFs for $\eta_c(1S)\to\phi\phi$ and $\omega\phi$~\cite{etac1S2ff}.
Projection of the best fit results onto the $\omega\omega$ (a), $\phi\phi$ (b) and $\omega\phi$ (c) mass.
}
\label{etac1S2ffwf}
\end{figure*}

In the $M1$ transition decay $\psi(2S)\to\gamma\eta_c(1S)$~\cite{etac1SBES3} , six hadronic decay modes ($K_S K^+\pi^-$, $K^+ K^-\pi^0$, $\eta\pi^+\pi^-$, $K_S K^+\pi^+\pi^-\pi^-$, $K^+ K^-\pi^+\pi^-\pi^0$, and $3(\pi^+\pi^-)$ (where the inclusion of charge conjugate mode is implied) are adopted to reconstruct $\eta_c(1S)$ meson.
The anomolous  line shape is seen around $\eta_c(1S)$, as illustrated in Fig.~\ref{etac1S}, which could be described well using  a combination of the energy-dependent hindered-M1 transition matrix element and a full interference with non-resonance $\psi(2S)$ radiative decays.
The measured mass and width of  $\eta_c(1S)$ , $(2984.3\pm0.6\pm0.6)$ MeV/$c^2$ and $(32.0\pm1.2\pm1.0)$ MeV, respectively, are in good agreement with those from
photon-photon fusion and $B$ decays~\cite{etac1Sbabar,etac1Sbelle,etac1Sbelle2}, which help clarify the discrepancies discussed above.
With the measured resonant parameters at BESIII, the hyperfine mass splitting is then determined to be $\Delta M_{\rm hf}(1S)$ $\equiv M(J/\psi(1S))-M(\eta_c(1S))$ $=112.6\pm0.8$ MeV/$c^2$, which agrees well with recent lattice computations~\cite{LQCD1,LQCD2,LQCD3} as well as quark-model predictions~\cite{qmpre}, and sheds light on spin-dependent interactions in quarkonium states. In addition, BESIII also reported the measurement of $\eta_c\to\omega\omega$ in the $M1$ transition decay $J/\psi\to\gamma\eta_c(1S)$~\cite{etac1S2ww}, as indicated in Fig.~\ref{etac1S2ffwf} (a), the measured mass and width are consistent with those from  $\psi(2S)\to\gamma\eta_c(1S)$~\cite{etac1SBES3}, while the uncertainties are larger due to the limited statistics.

\begin{figure}[hbtp]
\centering
\includegraphics[height=10cm,width=7.3cm]{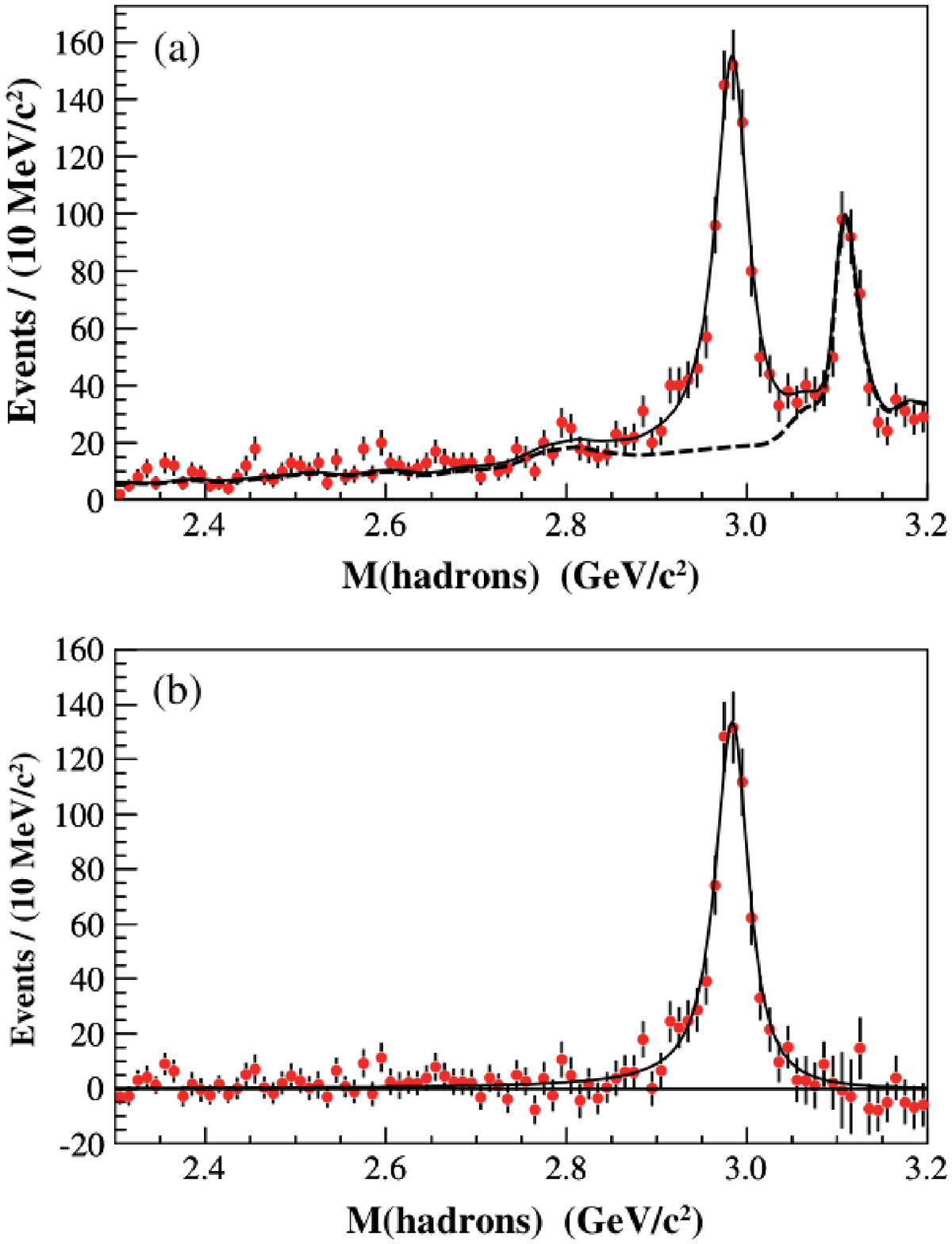}
\caption{The hadronic mass spectrum in $\psi(2S)\to\pi^0 h_c(1P)$, $h_c(1P)\to\gamma\eta_c(1S)$, $\eta_c(1S)\to X_i$ summed over the 16 final state $X_i$. The dots with error bars represent the hadronic
mass spectrum in data~\cite{etac1S216}. The solid line shows the total fit function and the dashed line is the background component of the fit. (b) The background-subtracted hadronic mass spectrum with the
signal shape overlaid. }
\label{etac1S216mode}
\end{figure}

Due to the high production rate of  $\eta_c(1S)$ in the $E1$ transition decay $h_c(1P)\to\gamma\eta_c(1S)$, the $\psi(2S)\to\pi^0 h_c(1P)$~\cite{etac1S216} decay also offers a unique place to investigate the $\eta_c(1S)$ properties, where the $\eta_c(1S)$ could be reconstructed with its decays to fully hadronic final states.  With sixteen exclusive hadronic decays,
BESIII reported the study of $\psi(2S)\to\pi^0 h_c(1P)$ with  $h_c(1P)\to\gamma\eta_c(1S)$~\cite{etac1S216}.
Of interesting is the $\eta_c(1S)$ line shape observed in $h_c(1P)\to\gamma\eta_c(1S)$, as displayed in Fig.~\ref{etac1S216mode}, seems contrary to those observed in both $J/\psi$ and $\psi(2S)$ radiative decays. And the mass and width are determined to be $M(\eta_c(1S))$ $=(2984.49\pm1.16\pm0.52)$ MeV/$c^2$ and $\Gamma(\eta_c(1S))$ $=(36.4\pm3.2\pm1.7)$ MeV, respectively, which are consistent with those via study the $\psi(2S)$ radiative decay $\psi(2S)\to\gamma\eta_c(1S)$~\cite{etac1SBES3} and $B$-factory results from $\gamma\gamma\to\eta_c(1S)$ and $B$ decays~\cite{etac1SBelle,etac1SBabar}. This consistence indicates that there is large interference amplitude in the $E1$ transitions of $\psi(2S)\to\gamma\eta_c(1S)$ and $J/\psi(1S)\to\gamma\eta_c(1S)$, and it is reasonable to take it into account when measuring the resonance parameters of $\eta_c(1S)$ within these decays.



With weighted least squares method~\cite{weighted}, and combining statistical and systematic errors in quadrature, the averaged mass for $\eta_c(1S)$ at BESIII is $(2984.68\pm0.93)$ MeV/$c^2$, and the mass splitting with $S$ wave iso-spin triplet is
$\Delta M_{\rm hf}(1S)$ $=(112.22\pm 0.93)$ MeV/$c^2$ which agrees well with recent lattice computations~\cite{LQCD1,LQCD2,LQCD3} as well as quark-model predictions~\cite{qmpre}, and sheds light on spin-dependent interactions in quarkonium states.

\subsection{$\eta_c(1S)$ decays}
\subsubsection{Decays into vector meson pairs}
The processes $\eta_c(1S)$ decaying into vector meson (abbreviated as $V$ hereafter) pairs are highly suppressed by the helicity selection rule (HSR)~\cite{hsr}, but experiment gives extremely large results~\cite{etac1S2VVBESB}, in which the BRs of $\eta_c(1S)\to VV$ are measured to be ${\cal{B}}(\eta_c(1S)\to\rho\rho)=(1.23\pm0.37\pm0.50)\times10^{-2}$, ${\cal{B}}(\eta_c(1S)\to K^*\overline{K}^*)=(10.3\pm2.6\pm4.3)\times10^{-3}$, ${\cal{B}}(\eta_c(1S)\to\phi\phi)=(2.5\pm0.5\pm0.9)\times10^{-3}$, ${\cal{B}}(\eta_c(1S)\to\omega\omega)<6.3\times10^{-3}$ (90\% C.L.), and
${\cal{B}}(\eta_c(1S)\to\omega\phi)<(1.7\times10^{-3}$ (90\% C.L.), respectively. This large discrepancy between theoretical predictions and experimental measurements stands as a long-term puzzle existing in the charmonium physics. The higher order radiative corrections fail to solve this issue, since they are all suppressed by the light quark masses. Although beyond the scope of pQCD some non-perturbative models have been put forward and considered to be solutions to the
problem, such as the intermediate meson exchange model~\cite{zhaoqpreetac}, the charmonium light Fock component admixture model~\cite{zhaoqpreetac2}, the $^3P_0$ quark pair creation mechanism
~\cite{preEtac2VV1}, and long-distance intermediate meson loop effects~\cite{preEtac2VVLiuQ}.

Predictions based on next-to-leading order (NLO) PQCD calculations and the so-called higher-twist contributions are also performed~\cite{preEtac2VV2,preEtac2VV3,twist}. The latter contribution is found to be dominate and give out a reasonably large decay width ($\sim10^{-4}$), though it still deviates a lot from the experimental measurement ($\sim 10^{-3}$).
To help understand the $\eta_c(1S)$ decay mechanism, high precision measurements of these BRs are desirable.

$\eta_c(1S)\to\omega\omega$, $\omega\phi$ and $\phi\phi$ are investigated via $J/\psi(1S)\to\gamma\eta_c(1S)$ at BESIII~\cite{etac1S2ww}~\cite{etac1S2ff}. Clear signal of $\eta_c(1S)\to\omega\omega$ are observed for the first time. A shown in Fig.~\ref{etac1S2ffwf} (a), by means of a partial wave analysis, the corresponding BF is measured to be ${\cal{B}}(\eta_c(1S)\to\omega\omega)$ $=(2.88\pm0.10\pm0.46\pm0.68)\times10^{-3}$, where the external uncertainty refers to that arising from the BF of the decay $J/\psi\to\gamma\eta_c(1S)$. Fig.~\ref{etac1S2ffwf} (b) and (c) show the invariant masses of $\phi\phi$ and $\omega\phi$. BR of $\eta_c(1S)\to\phi\phi$, $(2.5\pm0.3^{+0.3}_{-0.7}\pm0.6)\times10^{-3}$,  is consistent with the previous measurements~\cite{etac1S2VVBESB}, but the precision is improved. No significant signal for $\eta_c(1S)\to\omega\phi$ is observed. The upper limit at 90\% C.L. on the BR is determined to be ${\cal{B}}(\eta_c(1S)\to\omega\phi)$ $<2.5\times10^{-4}$, which is one order magnitude more stringent than the previous upper limit~\cite{etac1S2VVBESB}.

To understand the HSR violation mechanism, a comparison between the experimental measurements and the theoretical predictions based on the light quarkmass correction~\cite{preEtac2VV1},
 the $^3P_0$ quark pair creation mechanism~\cite{preEtac2VV2} and the intermediate meson loop effects~\cite{preEtac2VVLiuQ} is presented in Table~\ref{tab::BRetac2VV}.
We can find that the measured ${\cal{B}}(\eta_c(1S)\to\omega\omega)$ and ${\cal{B}}(\eta_c(1S)\to\phi\phi)$ are close to the predictions of the $^3P_0$ quark model~\cite{preEtac2VV2} and
the meson loop effects~\cite{preEtac2VVLiuQ}. In addition, the measured upper limit for ${\cal{B}}(\eta_c(1S)\to\omega\phi)$ is comparable with the predicted on in ~\cite{preEtac2VVLiuQ}.
 The consistency between data and the theoretical calculation indicates the importance of QCD higher twist
contributions or the presence of a NPQCD mechanism.

\begin{table*}[!hbtp]
\begin{center}
\caption{Comparison of BESIII measurement of BRs for $\eta_c(1S)\to\omega\omega,\phi\phi$ and $\omega\phi$ with the previous results and theoretical predictions, where `$-$' denotes there is no prediction.}
\begin{tabular}{lcccc}
\hline
   &    & ${\cal{B}}(\eta_c(1S)\to\omega\omega)(\times10^{-3})$ & ${\cal{B}}(\eta_c(1S)\to\phi\phi)(\times10^{-3})$ & ${\cal{B}}(\eta_c(1S)\to\omega\phi)(\times10^{-3})$\\
\hline
\multirow{2}*{Exp. results}   & BESIII~\cite{etac1S2ww}~\cite{etac1S2ff}             & $2.88\pm0.10\pm0.46\pm0.68$    & $2.5\pm0.3^{+0.3}_{-0.7}\pm0.6$  & $<0.25$   \\
                                      & BESII~\cite{etac1S2VVBESB}              & $<6.3\times10^{-3}$            & $1.9\pm0.6$                      & $<1.7$    \\
                                      &                                      &&&\\
\multirow{3}*{The. predictions}  &  PQCD~\cite{preEtac2VV1}                   & 0.09$\sim$0.13     & 0.7$\sim$0.8 & $-$\\
                                        &  $^3P_0$ quark model~\cite{preEtac2VV2}    & 1.5$\sim$1.6       & 1.9$\sim$2.0 & 0\\
                                        & Charm meson loop~\cite{preEtac2VVLiuQ}        & 1.8           & 2.0     & $<0.33$\\
\hline
\end{tabular}
\label{tab::BRetac2VV}
\end{center}
\end{table*}

\subsubsection{Decays into baryon pairs}
$\eta_c(1S)$ decaying to baryon-anti-baryon pairs are also supposed to be highly suppressed by the HSR as a consequence of the PQCD framework. The contradictions between PQCD calculations and experimental measurements also exist in this process, as is the discrepancy for $\eta_c(1S)\to VV$. Many theoretical models have been developed to understand these contradictions, such as by the quark-diquark model for the proton~\cite{VHSR1,VHSR2},
constituent quark-mass corrections~\cite{VHSR3,VHSR4}, mixing between the charmonium state and the glueball~\cite{VHSR5}, and the quark pair creation model~\cite{VHSR6}. However, none of them is perfect enough.

In Refs.~\cite{zhaoq1,zhaoq2}, a theory related to intermediate meson loop (IML) transitions is proposed, where the long-distance interaction can evade the OZI rule and allow the violation of the PQCD HSR.
 Further calculations on the BFs of $\eta_c(1S)\to B_8 \bar{B}_8$, $\chi_{c0}(1P)\to B_8\bar{B}_8$ and $h_c(1P)\to B_8\bar{B}_8$, where $B_8$ denote the octet baryon,
based on charmed-meson loops were carried out~\cite{zhaoq3}, and the results agree with the measured BFs of
$\eta_c(1S)\to p\bar{p}$ and $\eta_c(1S)\to\Lambda\bar\Lambda$, but with large uncertainty~\cite{etac1S2BBbelle}~\cite{etac1S2LL}.

\begin{figure*}[hbtp]
\centering
\includegraphics[height=4cm,width=5.3cm]{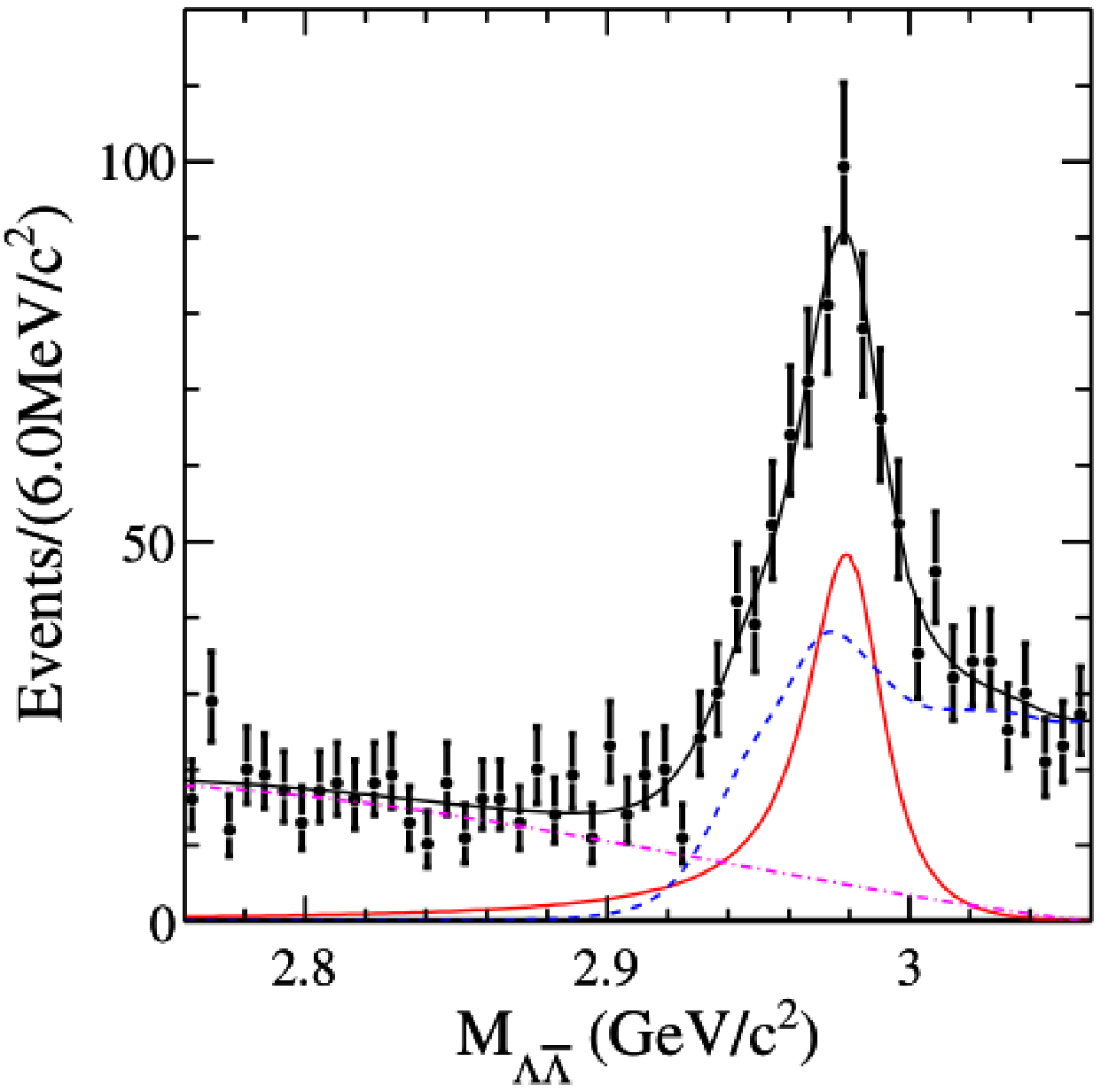}
\put(-70,100){\bf(a)}
\includegraphics[height=4cm,width=5.3cm]{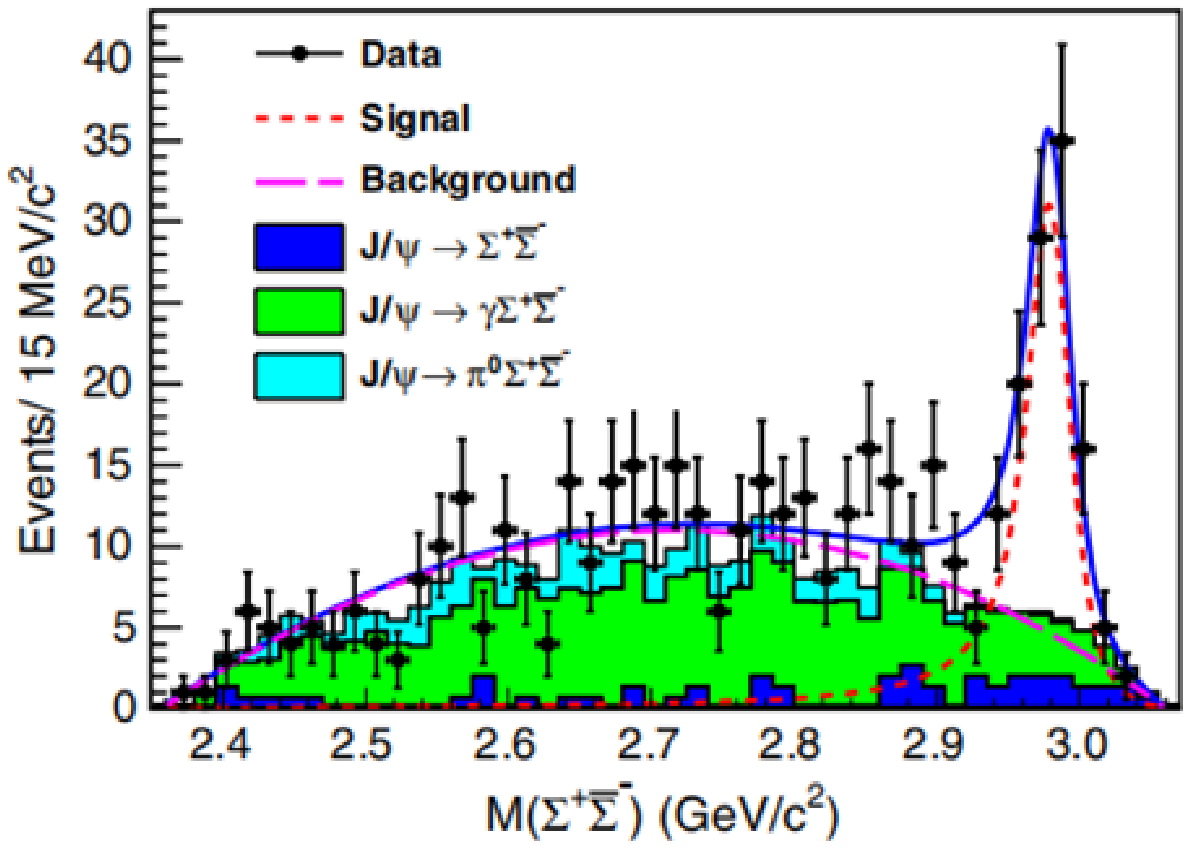}
\put(-70,100){\bf(b)}
\includegraphics[height=4cm,width=5.3cm]{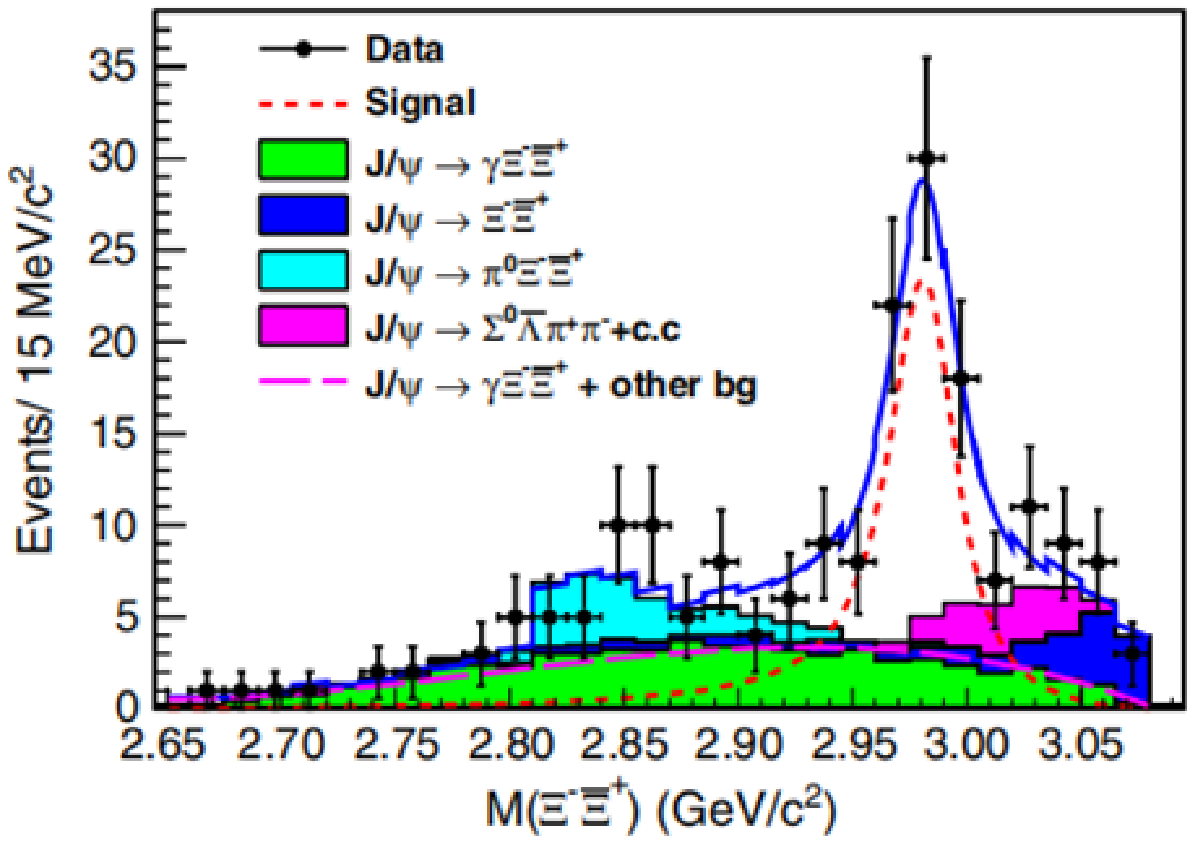}
\put(-70,100){\bf(c)}
\caption{Observation of $\eta_c(1S)\to\Lambda\bar\Lambda$~\cite{etac1S2LL}, $\Sigma^+\bar{\Sigma}^-$ and $\Xi^-\bar\Xi^+$~\cite{etac1S2BBbar}.
Invariant-mass distributions of $\Lambda\bar\Lambda$ (a), $\Sigma^+\bar{\Sigma}^-$ (b), and $\Xi^-\bar\Xi^+$ (c), as well as the fitted curves.
Dots with error bars are data, and the histograms are the backgrounds from simulated $J/\psi$ decays.
Solid lines are the total fit results, signals are shown in short-dashed lines, and backgrounds are shown as long-dashed lines and shaded
histograms.
}
\label{etac1S2BB}
\end{figure*}

BESIII Collaboration reported a series of results on $\eta_c(1S)\to B_8\bar{B}_8$ via $\jpsi(1S)$ radiative transition $\jpsi(1S)\to\gamma\eta_c(1S)$ for the first time. e.g., $\eta_c(1S)\to\Lambda\overline\Lambda$, $\Sigma^+\bar\Sigma^-$ and $\Xi^-\bar\Xi^+$~\cite{etac1S2LL}~\cite{etac1S2BBbar}, as well as the decay $\eta_c(1S)\to p\overline{p}$ via $\psi(2S)\to\pi^0 h_c(1P)$~\cite{etac1S216} and $e^+e^-\to\pi^+\pi^-h_c(1P)$~\cite{etac1S2mxn}, respectively (see detail for $\eta_c(1S)\to p\overline{p}$ in Sec.~\ref{sec:etac2LH}). Figure~\ref{etac1S2BB} (a), (b) and (c) show the invariant mass spectrum for $\Lambda\overline\Lambda$, $\Sigma^+\bar\Sigma^-$ and $\Xi^-\bar\Xi^+$ final states, respectively. The corresponding measured BRs are listed in Tab.~\ref{tab::etac2BB}.
The decay $\eta_c(1S)\to\Sigma^+\bar\Sigma^-$ and $\Xi^-\bar\Xi^+$ are observed for the first time, and BR for $\eta_c(1S)\to p\overline{p}$ agrees with the previous measurements, but with higher precision~\cite{etac1S2BBbelle}.

Table.~\ref{tab::etac2BB} compares the results of BESIII measurement of BRs for $\eta_c(1S)\to B_8\overline{B}_8$ with the previous results and theoretical predictions by the intermediate charmed hadron loop transitions~\cite{zhaoq3}. The measured BF of $\eta_c(1S)\to\Sigma^+\bar\Sigma^-$ is larger than the predictions from charmed-meson loop calculations~\cite{zhaoq3}, while the measured BF of $\eta_c(1S)\to\Xi^-\bar\Xi^+$ agrees with the prediction. Among the four $\eta_c(1S)$ baryonic decays ($\eta_c(1S)\to p\bar p$, $\Lambda\bar\Lambda$, $\Sigma^+\bar\Sigma^-$, and $\Xi^-\bar\Xi^+$),
only $\eta_c(1S)\to\Sigma^+\bar\Sigma^-$ disagrees with the prediction, which may indicate the violation of SU(3) symmetry~\cite{etac1S2BBbar}.

\begin{table*}[hbtp]
\begin{center}
\caption{Comparison of BESIII measurement of BRs for $\eta_c(1S)\to B_8\overline{B}_8$ with the previous results and theoretical predictions by the intermediate charmed hadron loop transitions~\cite{zhaoq3}.
Here, the third uncertainty 
is from input branching fractions taken from Ref.~\cite{pdg}.
}
\begin{tabular}{lcccc}
\hline
  &  ${\cal{B}}(\eta_c(1S)\to p\overline{p})(\times10^{-4})$  & ${\cal{B}}(\eta_c(1S)\to \Lambda\overline{\Lambda})(\times10^{-4})$ &  ${\cal{B}}(\eta_c(1S)\to\Sigma^-\overline{\Sigma}^+)(\times10^{-4})$  & ${\cal{B}}(\eta_c(1S)\to\Xi^-\overline{\Xi}^+)(\times10^{-4})$\\
\hline

\multirow{2}*{Exp. results}   &  $12.0\pm2.6\pm1.5$~\cite{etac1S2mxn} & $11.6\pm1.2\pm1.9\pm2.8$~\cite{etac1S2LL} & \multirow{2}*{$21.1\pm2.8\pm1.8\pm5.0$~\cite{etac1S2BBbar}}   & \multirow{2}*{$8.9\pm1.6\pm0.8\pm2.1$~\cite{etac1S2BBbar}}\\
                                     & $15.8\pm1.2^{+1.8}_{-2.2}\pm4.7$~\cite{etac1S2BBbelle} & $8.7^{+2.4+0.9}_{-2.1-1.4}\pm2.7$~\cite{etac1S2BBbelle} &&\\
                                     &&&&\\
\multirow{2}*{\shortstack{The. predictions\\(Hadron loop)}}  &  \multirow{2}*{$9.0\sim17.0$}   & \multirow{2}*{$6.3\sim12.5$}    &   \multirow{2}*{$5.05\sim10.0$}   & \multirow{2}*{$4.82\sim9.56$}\\
&&&&\\
\hline
\end{tabular}
\label{tab::etac2BB}
\end{center}
\end{table*}

\subsubsection{Decay into light hadrons}\label{sec:etac2LH}
As mentioned above, the $\eta_c(1S)$ decays to light hadron final states has been investigated with $h_c(1P)\to\gamma\eta_c(1S)$ via $\psi(2S)\to\pi^0 h_c(1P)$. In addition, BESIII detector has also collected sizable data samples between 4.009 and 4.600 GeV (called ``{\sc XYZ} data" hereafter) since 2013 to study the {\sc XYZ} states~\cite{xyzdata2015}. A large production rate of $e^+e^-\to\pi^+\pi^-h_c(1P)$ has been found~\cite{Zc40420}.  The total number of $h_c(1P)$ events in all these data samples combined is comparable to that from $\psi(2S)\to\pi^0 h_c(1P)$ according to the measured cross section and the corresponding integrated luminosity at each energy point. The $h_c(1P)$ is tagged by the recoil mass ($RM$) of $\pi^+\pi^-$ in {\sc XYZ} data. Compared with that tagged by $RM(\pi^0)$ in $\psi(2S)$ data sample, the former has lower background and higher detection efficiency than the latter due to the better momentum resolution for charged track. Using the data at $\sqrt{s}=$ 4.23,
4.26, 4.36 and 4.42 GeV, the BF of four $\eta_c$ exclusive decays are measured via the process $e^+e^-\to\pi^+\pi^-h_c(1P)$, $h_c(1P)\to\gamma\eta_c(1S)$~\cite{etac1S2mxn}. These exclusive decays are $\eta_c(1S)\to K^+K^-\pi^0$, $K^0_S K^\pm\pi^\mp$, $2(\pi^+\pi^-\pi^0)$ and $p\bar{p}$, respectively. The BFs of $\eta_c(1S)$ exclusive decays are obtained by a simultaneous fit to the $RM(\pi^+\pi^-\gamma)$ for both inclusive and exclusive modes. As a result, we obtained the highest precision compared to the previous single measurement.
The measured BFs are summarized in Tab.~\ref{tab::etachadrons}. A few more discussion can be found in Sec.~\ref{sec:hc2LH}.

\begin{table*}[!hbtp]
\begin{center}
\caption{Review of the measured BF of $\eta_c(1S)\to X_i$ at BESIII.
${\cal{B}}_{\psi(2S)}$ denote the BFs measured with $\psi(2S)\to\pi^0 h_c(1P)$ and $h_c(1P)\to\gamma\eta_c(1S)$,
${\cal{B}}_{e^+e^-}$ denote the BFs measured with $e^+e^-\to\pi^+\pi^- h_c(1P)$ and $h_c(1P)\to\gamma\eta_c(1S)$.
Here, $X_i$ denotes hadronic final states. For the measured BFs, the first uncertainties are statistical, second ones are systematic, third ones are uncertainties due to ${\cal{B}}(\psi(2S)\to\pi^0 h_c(1P))$ and ${\cal{B}}(h_c(1P)\to\gamma\eta_c(1S))$. The last column gives the status of ${\cal{B}}^{\rm PDG}_{\eta_c(1S)\to X_i}$ before the BESIII experiment.}
\begin{tabular}{lccc}
\hline
Decay mode &  ${\cal{B}}_{\psi(2S)}(10^{-2})$~\cite{etac1S216}  & ${\cal{B}}_{e^+e^-}(10^{-2})$~\cite{etac1S2mxn} &    Comment$(10^{-2})$\\
\hline
$\eta_c(1S)\to p\bar{p}$                          &  $0.15\pm0.04\pm0.02\pm0.01$ & $0.120\pm0.026\pm0.015$       &    $0.141\pm0.017$\\
$\eta_c(1S)\to 2(\pi^+\pi^-)$                     &  $1.72\pm0.19\pm0.25\pm0.17$ & ...                  &    $0.86\pm0.13$\\
$\eta_c(1S)\to K^+K^-K^+K^-$                      &  $0.22\pm0.08\pm0.03\pm0.02$ & ...                  &    $0.134\pm0.032$\\
$\eta_c(1S)\to K^+K^-\pi^+\pi^-$                  &  $0.95\pm0.17\pm0.13\pm0.09$ & ...                  &    $0.61\pm0.12$\\
$\eta_c(1S)\to p\bar{p}\pi^+\pi^-$                &  $0.53\pm0.15\pm0.08\pm0.05$ & ...                  &    $<1.2$ (at 90\% C.L.)\\
$\eta_c(1S)\to 3(\pi^+\pi^-)$                     &  $2.02\pm0.36\pm0.36\pm0.19$ & ...                  &    $1.5\pm0.5$\\
$\eta_c(1S)\to K^+K^-2(\pi^+\pi^-)$               &  $0.83\pm0.39\pm0.15\pm0.08$ & ...                  &   $0.71\pm0.29$\\
$\eta_c(1S)\to K^+K^-\pi^0$                       &  $1.04\pm0.17\pm0.11\pm0.10$ & $1.15\pm0.12\pm0.10$ &   $1.2\pm0.1$\\
$\eta_c(1S)\to p\bar{p}\pi^0$                     &  $0.35\pm0.11\pm0.05\pm0.03$ & ...                  &   first measurement\\
$\eta_c(1S)\to K^0_S K^\pm\pi^\mp$                &  $2.60\pm0.29\pm0.34\pm0.25$ & $2.60\pm0.21\pm0.20$ &   $2.4\pm0.2$\\
$\eta_c(1S)\to K^0_S K^\pm\pi^\mp\pi^\pm\pi^\mp$  &  $2.75\pm0.51\pm0.47\pm0.27$ & ...                  &   first measurement\\
$\eta_c(1S)\to \pi^+\pi^-\eta$                    & $1.66\pm0.34\pm0.26\pm0.16$  & ...                  &   $4.9\pm1.8$\\
$\eta_c(1S)\to K^+K^-\eta$                        & $0.48\pm0.23\pm0.07\pm0.05$  & ...                  &   $<1.5$ (at 90\% C.L)\\
$\eta_c(1S)\to 2(\pi^+\pi^-)\eta$                 & $4.40\pm0.86\pm0.85\pm0.42$  & ...                  &   first measurement\\
$\eta_c(1S)\to \pi^+\pi^-\pi^0\pi^0$              & $4.66\pm0.50\pm0.76\pm0.45$  & ...                  &   first measurement\\
$\eta_c(1S)\to 2(\pi^+\pi^-\pi^0)$                & $17.23\pm1.70\pm2.29\pm1.66$ & $15.3\pm1.8\pm1.8$   &   first measurement\\
\hline
\end{tabular}
\label{tab::etachadrons}
\end{center}
\end{table*}

\begin{figure*}[hbtp]
\centering
\includegraphics[height=5cm,width=7.3cm]{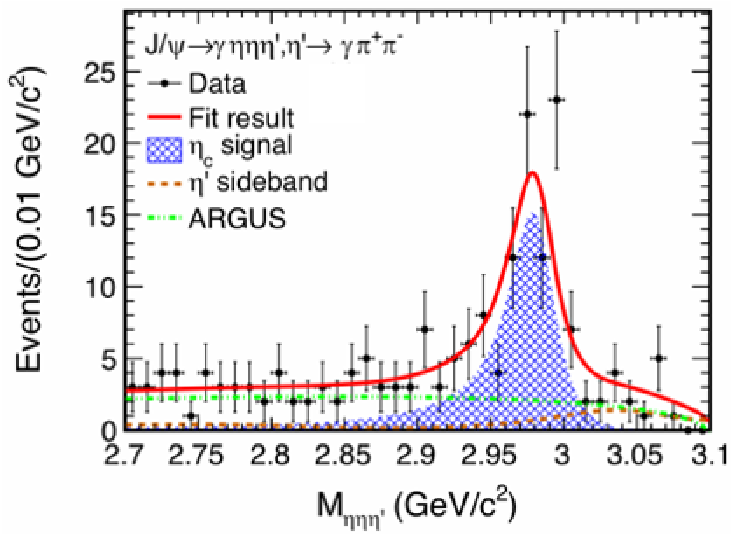}
\includegraphics[height=5cm,width=7.3cm]{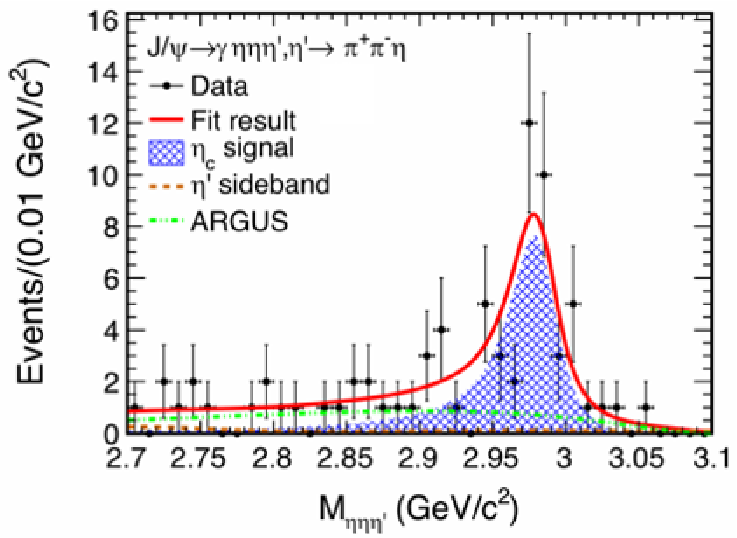}\\
\caption{Fit results for $\eta_c(1S)$ in the invariant mass distribution of $\eta\eta\eta'$ for the decays of $\eta'\to\gamma\pi^+\pi^-$ and $\eta'\to\eta\pi^+\pi^-$, respectively.~\cite{etac1S2etaetaetap}.}
\label{etac1S2etaetaetap}
\end{figure*}

The decay width of $\eta_c(1S)$ into a pseudoscalar glueball is computed in the framework
of a $U(4)_r\times U(4)_l$ symmetric linear sigma model with (pseudo-)scalar and (axial-) vector mesons~\cite{etac1S2etaetaetappre}.
Using the general formula for the three-body decay width for $\eta_c$~\cite{pdg}, the corresponding tree-level decay amplitudes for $\eta_c(1S)\to \eta\eta\eta'$ are obtained and the corresponding partial decay widths of $\eta_c(1S)\to \eta\eta\eta'$ are predicted to be $\Gamma(\eta_c(1S)\to\eta\eta\eta')$ $=0.045\pm0.014$ MeV. Experimentally, first observation of $\eta_c(1S)\to\eta\eta\eta'$ via
$J/\psi(1S)\to\gamma\eta\eta\eta'$ is reported at BESIII~\cite{etac1S2etaetaetap}. Figure~\ref{etac1S2etaetaetap} shows the fit results for $\eta_c(1S)$ in the invariant mass distribution of $\eta\eta\eta'$ with  $\eta'\to\gamma\pi^+\pi^-$ and $\eta'\to\eta\pi^+\pi^-$, respectively. A clear $\eta_c(1S)$ signal is seen, and the ${\cal{B}}(J/\psi(1S)\to\gamma\eta_c(1S))$$\times{\cal{B}}(\eta_c(1S)\to\eta\eta\eta')$ is determined to be $(4.86\pm0.62\pm0.45)\times10^{-5}$, which is compatible with the theoretical prediction of partial decay width of $\eta_c(1S)\to\eta\eta\eta'$ in Ref.~\cite{etac1S2etaetaetappre}.

\subsubsection{Decays into $\gamma\gamma$}
The quarkonium annihilation rates are used to evaluate the strong fine-structure constant $\alpha_{S}$.
The square of the wave function at the origin cancels out in the ratio of partial widths~\cite{etac1S2ggpre},
\begin{eqnarray}
\frac{\Gamma(\eta_c(1S)\to\gamma\gamma)}{\Gamma(J/\psi(1S)\to\mu^+\mu^-)}
=\frac{4}{3}\lbrack 1+1.96\frac{\alpha_{S}(m^2_c)}{\pi}\rbrack,\label{eq3}
\end{eqnarray}
Using the `evaluated' partial widths in Ref.~\cite{pdg}, $\Gamma(\eta_c(1S)\to\gamma\gamma)$
$=5.15\pm0.35$ keV and $\Gamma(J/\psi(1S)\to\mu^+\mu^-)$ $=5.54\pm0.17$ keV, one finds that
$(3/4)\Gamma(\eta_c(1S)\to\gamma\gamma)/\Gamma(J/\psi(1S)\to\mu^+\mu^-)$ $=0.93\pm0.07$, which is consistent
with Eq.~\ref{eq3} but not precise enough to test the QCD model. A more
precise test would have taken into account $m(J/\psi(1S))\neq2m_c$ and the running of $\alpha_{S}$.

The total width of $\eta_c(1S)$ is dominated by the $gg$ final state. Its value is $32.0\pm0.7$ MeV~\cite{pdg}.
By inputting the value of $\Gamma(\eta_c(1S)\to\gamma\gamma)$, one can determine the $gg/\gamma\gamma$ ratio to be
\begin{eqnarray}
\frac{\Gamma(\eta_c(1S)\to gg)}{\Gamma(\eta_c(1S)\to\gamma\gamma)}
=\frac{9[\alpha_S(m^2_c)]^2}{8\alpha^2}\lbrack 1+8.2\frac{\alpha_{S}(m^2_c)}{\pi}\rbrack,\label{eq4}
\end{eqnarray}
which induces to $\alpha_S(m^2_c)$. This value should be regarded with caution in view of the large QCD
correction factor $1+8.2\alpha_S/\pi\sim1.8$.

Evidence of the $\eta_c(1S)\to\gamma\gamma$ decay is reported at BESIII~\cite{etac1S2ggbes3}, and the branching fraction of
$J/\psi\to\gamma\eta_c(1S)$ and $\eta_c(1S)\to\gamma\gamma$ is determined to be
${\cal{B}}(J/\psi\to\gamma\eta_c(1S),\eta_c(1S)\to\gamma\gamma)$ $=(4.5\pm1.2\pm0.6)\times10^{-6}$,
which agrees with the result from two-photon fusion~\cite{pdg}. Most recently, it was point out that the decay width of $\eta_c(1S)\to\gamma\gamma$ could be understood within
the $c\bar{c}$-glueball framework~\cite{etac1S2ggpre2}, but the discrepancy between the theoretical predictions and the experimental results of the partial width of $J/\psi\to\gamma\eta_c(1S)$ cannot be alleviated yet, and further precision measurement is needed.

\subsubsection{Decay into $CP$ and isospin violation modes}
\begin{figure*}[hbtp]
\centering
\includegraphics[height=5cm,width=7.3cm]{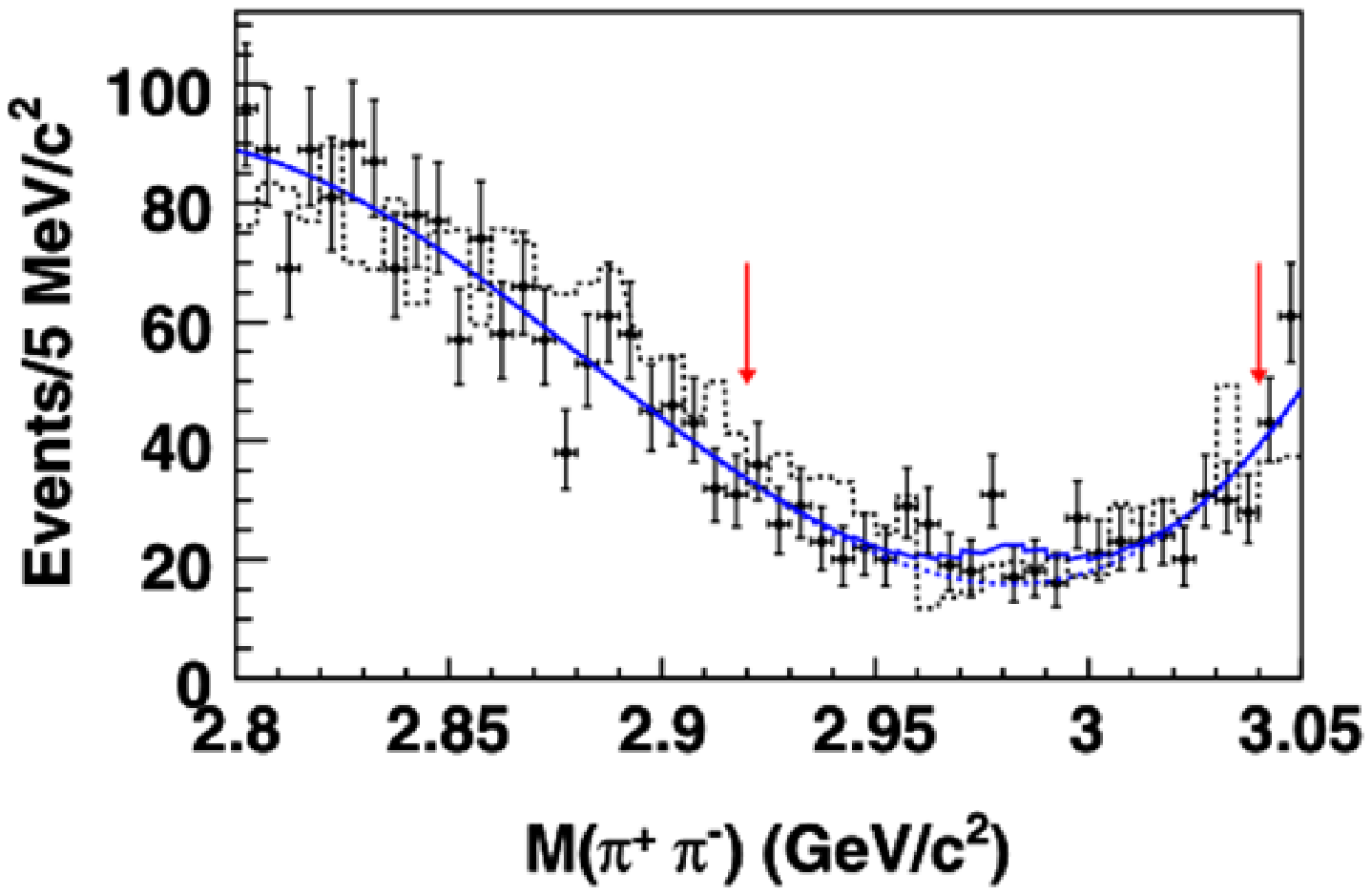}
\includegraphics[height=5cm,width=7.3cm]{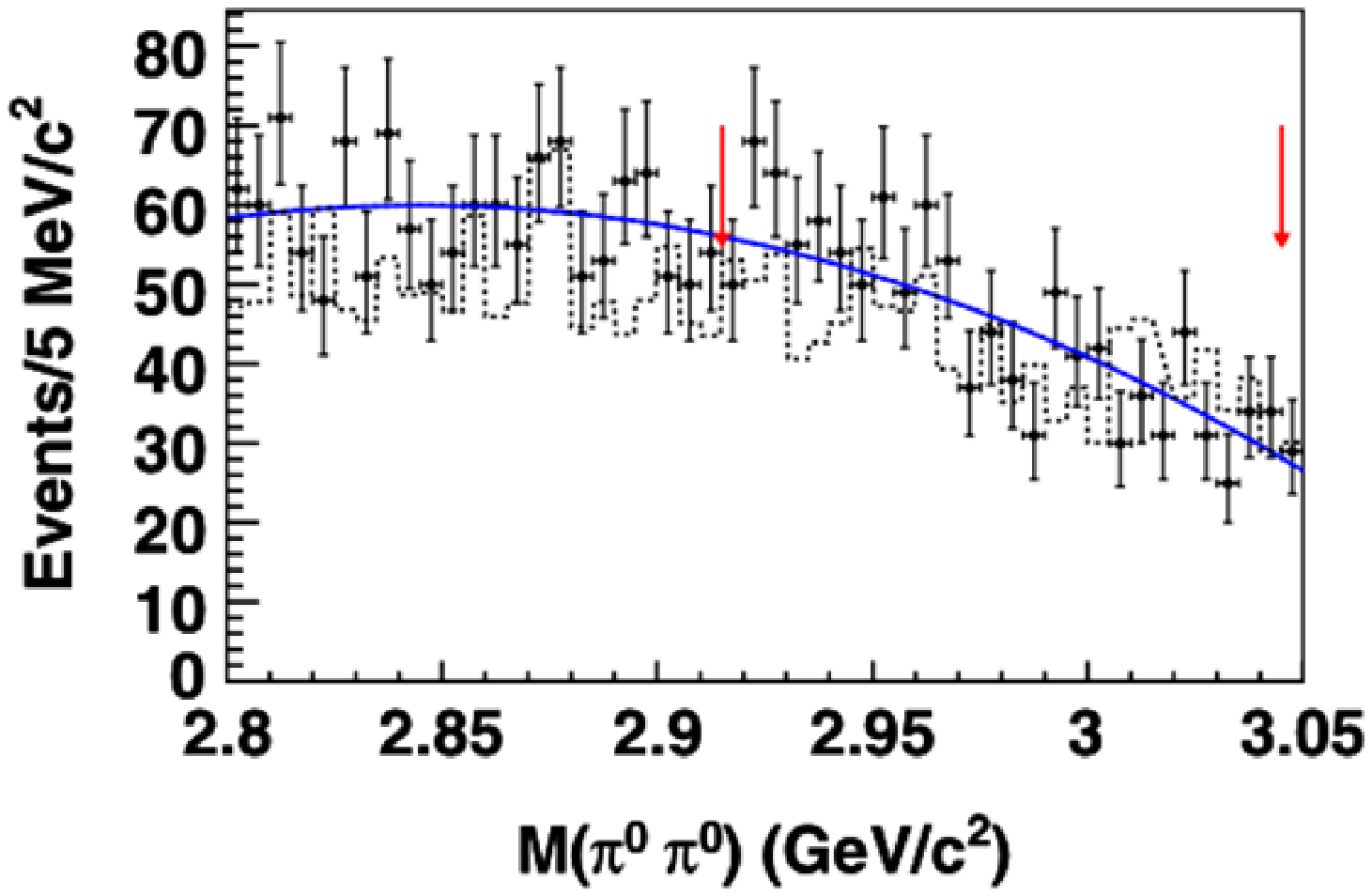}\\
\caption{Search for $CP$ and $P$ violating pseudoscalar decays into $\pi\pi$. The $\pi^+\pi^-$ and $\pi^0\pi^0$ invariant mass distributions of the final candidate events in the $\eta_c(1S)$ mass region.
 The dots with error bars are data, the solid lines are the best fit, and the dashed histograms are the sum of
all the simulated normalized backgrounds. The arrows show mass regions which contain around 95\% of the signal according to MC
simulations~\cite{etac1S2PP}.
}
\label{etac1S2PiPi}
\end{figure*}

$CP$ symmetry violation has important consequences; it is one of the key
ingredients for the matter-antimatter asymmetry in our universe. $CP$ violation can be experimentally searched for in processes such as meson decays.
The decays $\eta_c(1S)\to\pi^+\pi^-$ and $\pi^0\pi^0$, which violate both $P$ and $CP$ invariance, provide an excellent laboratory for testing the validity of symmetries of the physical world. In the SM, such decays can proceed only via the weak interaction with a BF of order $10^{-27}$, according to Ref.~\cite{etac2pipiPre}. Higher BFs are possible
either by introducing a CP violating term in the QCD Lagrangian (a BR up to $10^{-17}$ can be obtained in this scheme) of allowing $CP$ violation in the extended Higgs sector (with BF up to $10^{-15}$), as described in Ref.~\cite{etac2pipiPre}. The detection of these decays at any level accessible today would be the signal $P$ and $CP$ violations from new sources, beyond any considered extension of the SM. The available results for $\eta_c(1S)\to\pi^+\pi^-$ and $\pi^0\pi^0$ are ${\cal{B}}(\eta_c(1S)\to\pi^+\pi^-)$ $<6\times10^{-4}$ and ${\cal{B}}(\eta_c(1S)\to\pi^0\pi^0)$ $<4\times10^{-4}$ at the 90\%
C.L.~\cite{etac2pipiPreBES}, respectively. Searching for these decays directly via $J/\psi(1S)\to\gamma\eta_c(1S)$ at BESIII based on the first round $J/\psi(1S)$ data sample are preformed~\cite{etac1S2PP}. The final $\pi^+\pi^-$ and $\pi^0\pi^0$ mass spectra, as well as the fit result are shown in Fig.~\ref{etac1S2PiPi}. No significant $\eta_c(1S)$ signal is observed. Using the Bayesian method, the 90\% C.L. upper limits are determined to be ${\cal{B}}(\eta_c(1S)\to\pi^+\pi^-)$ $<1.3\times10^{-4}$ and ${\cal{B}}(\eta_c(1S)\to\pi^0\pi^0)$ $<4.2\times10^{-5}$. These results provide experimental limits for theoretical models, and predict how much $CP$ and $P$ violation may be observed in $\eta_c(1S)$ meson decays.

\begin{figure}[hbtp]
\centering
\includegraphics[height=5cm,width=7.3cm]{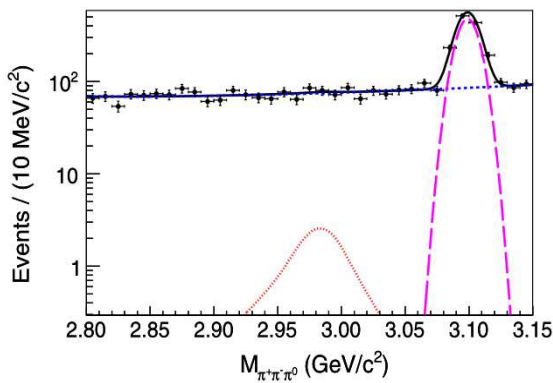}
\caption{Search for the isospin violation decay $\eta_c(1S)\to\pi^+\pi^-\pi^0$ via $\psi(2S)\to\gamma\eta_c(1S)$. The $\pi^+\pi^-\pi^0$ invariant mass distributions of the final candidate events in the $\eta_c(1S)$ mass region.
 The dots with error bars are data, the black solid line is the best fit, the red dashed line is $\eta_c(1S)$ signal, the long-dashed curve is the $J/\psi(1S)$
 background, and the short-dashed curve is the main background~\cite{etac1S23Pi}.
}
\label{etac1S23Pi}
\end{figure}

Search for the isospin violating mode $\eta_c(1S)\to\pi^+\pi^-\pi^0$ is performed for the first time~\cite{etac1S23Pi} at BESIII using two rounds $\psi(2S)$ data sample. The final $\pi^+\pi^-\pi^0$ mass spectra, as well as the fit result are shown in Figure~\ref{etac1S23Pi}. No obvious $\eta_c(1S)$ signal is seen, and the upper limit on ${\cal{B}}(\psi(2S)\to\gamma\eta_c(1S))$ $\times$ ${\cal{B}}(\eta_c(1S)\to\pi^+\pi^-\pi^0)$ is $1.6\times10^{-6}$ is given at 90\% C.L..
Using the BF of $\psi(2S)\to\gamma\eta_c(1S)$, $[3.4\pm0.5]\times10^{-3}$, the upper limit for ${\cal{B}}(\eta_c(1S)\to\pi^+\pi^-\pi^0)$
is calculated to be $5.5\times10^{-4}$. This is important to test isospin symmetry~\cite{isospin1,isospin2}.

\section{$\eta_c(2S)$ physics}
\label{sec:etac2Sdecays}
The production of the $\eta_c(2S)$ through a radiative transition from the $\psi(2S)$ involves a charmed-quark spin-flip and, thus, proceeds via a M1 transition. The BF has been predicted in the range ${\cal{B}}(\psi(2S)\to\gamma\eta_c(2S))$
$=(0.1-6.2)\times10^{-4}$~\cite{preetac2S1,preetac2S2,preetac2S3}.
A phenomenological prediction, by assuming that the matrix element governed $\psi(2S)\to\gamma\eta_c(2S)$ is the same as that for $J/\psi\to\gamma\eta_c(1S)$, is given by
\begin{eqnarray}
{\cal{B}}(\psi(2S)\to\gamma\eta_c(2S))
=\frac{k^3_{\psi(2S)}}{k^3_{J/\psi}}\frac{\Gamma_{J/\psi}}{\Gamma_{\psi(2S)}}{\cal{B}}(J/\psi\to\gamma\eta_c(1S)),\label{eq3-1}
\end{eqnarray}
where $k_{\psi(2S)}$ ($k_{J/\psi}$) is the photon energy for the $\psi(2S)\to\gamma\eta_c(2S)$ ($J/\psi\to\gamma\eta_c(1S)$) transition, $\Gamma_{\psi(2S)}$ ($\Gamma_{J/\psi}$)
is the $\psi(2S)$ ($J/\psi$) full width, and ${\cal{B}}(J/\psi\to\gamma\eta_c)$ $=(1.7\pm0.4)$\%~\cite{pdg}. Using the PDG values for $k_{\psi(2S)}$, $k_{J/\psi}$, $\Gamma_{\psi(2S)}$ and
$\Gamma_{J/\psi}$ leads to a prediction of $\psi(2S)\to\gamma\eta_c(2S)$ $=(3.9\pm1.1)\times10^{-4}$~\cite{searchEtac2SCLEO}.


\subsection{First observation of $\eta_c(2S)$ signal in $\psi(2S)$ M1 transition}
\begin{figure*}[hbtp]
\centering
\includegraphics[height=4cm,width=5.3cm]{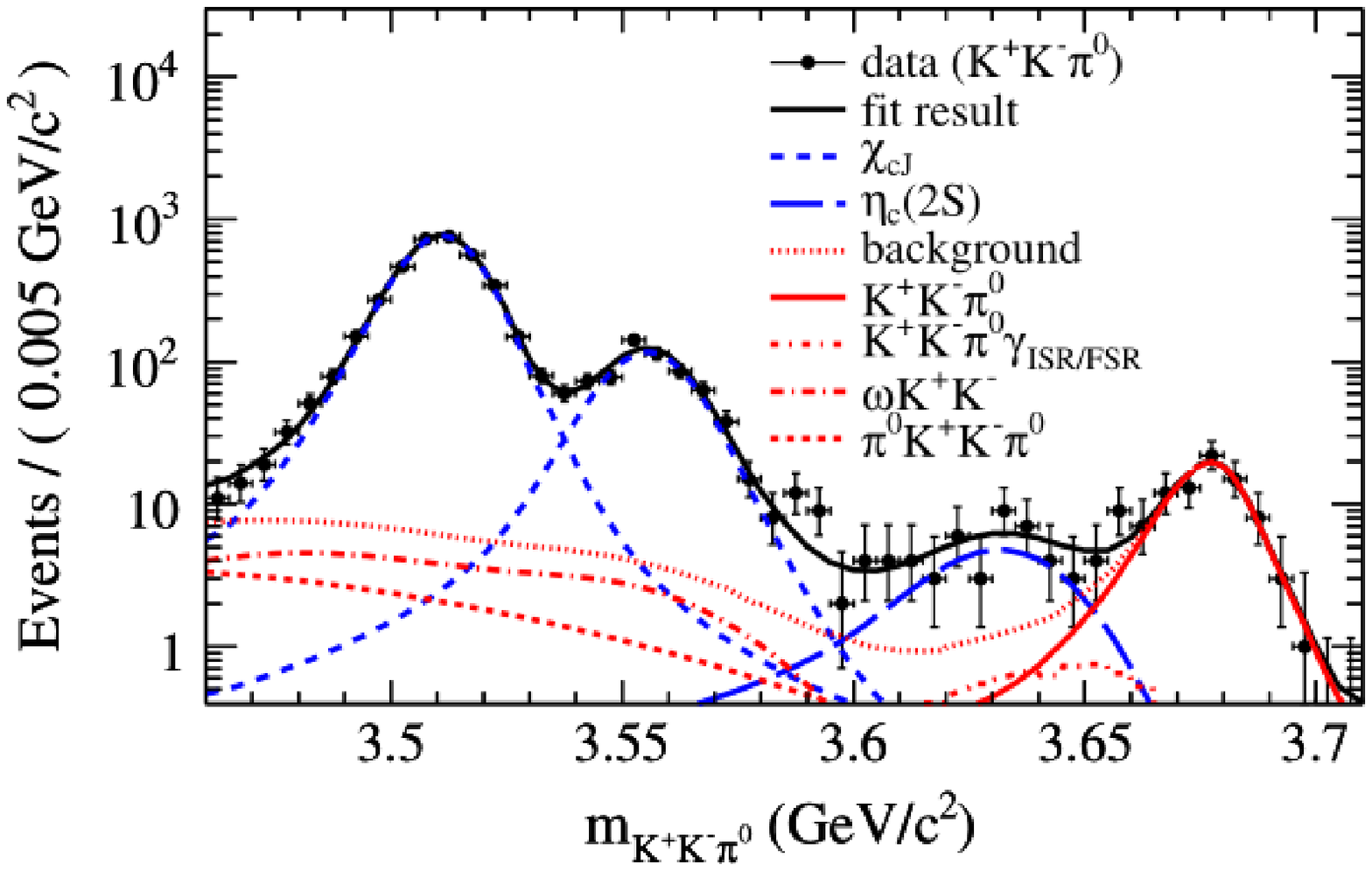}
\put(-90,100){\bf(a)}
\includegraphics[height=4cm,width=5.3cm]{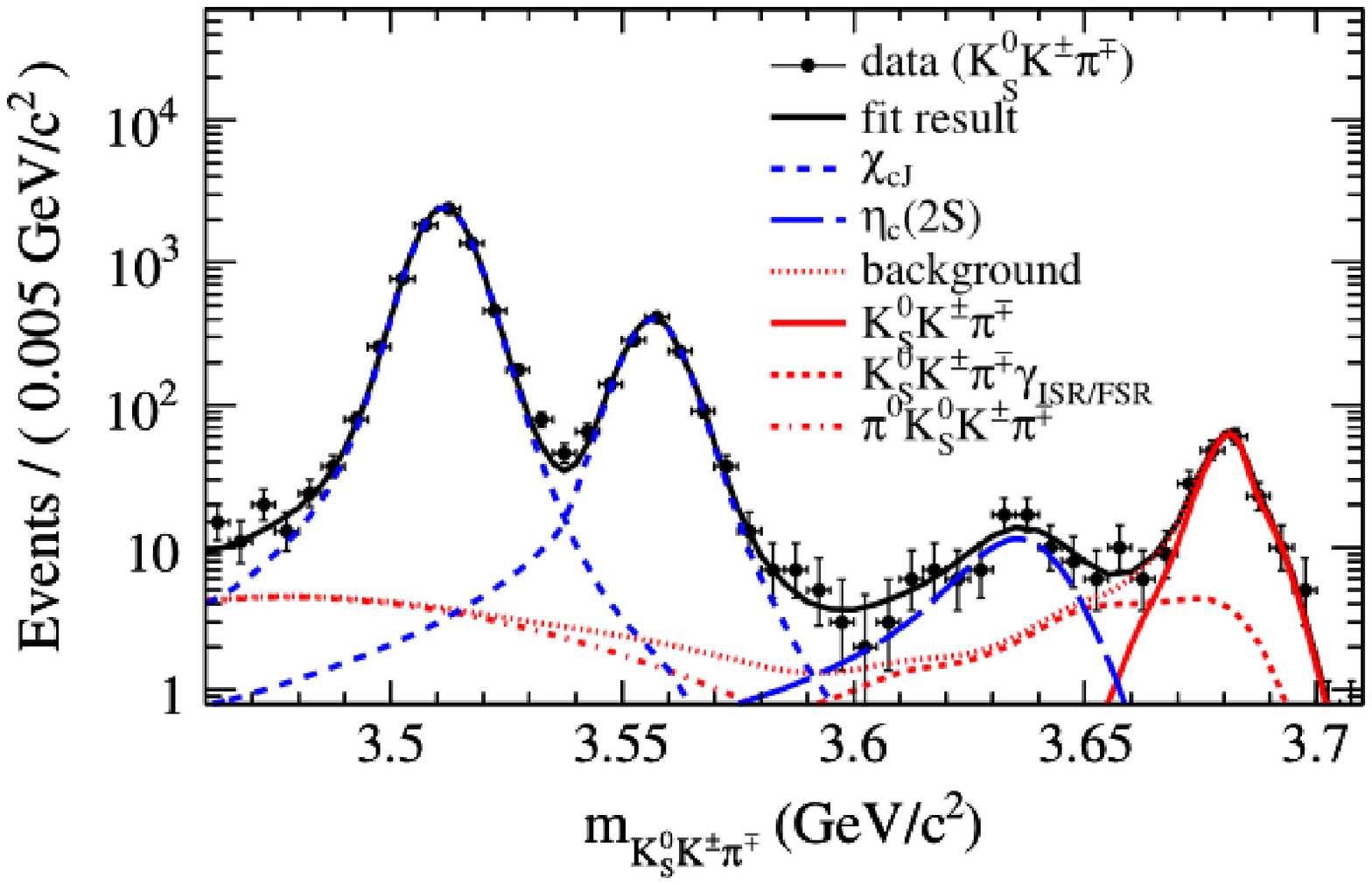}
\put(-90,100){\bf(b)}\\
\includegraphics[height=4cm,width=5.3cm]{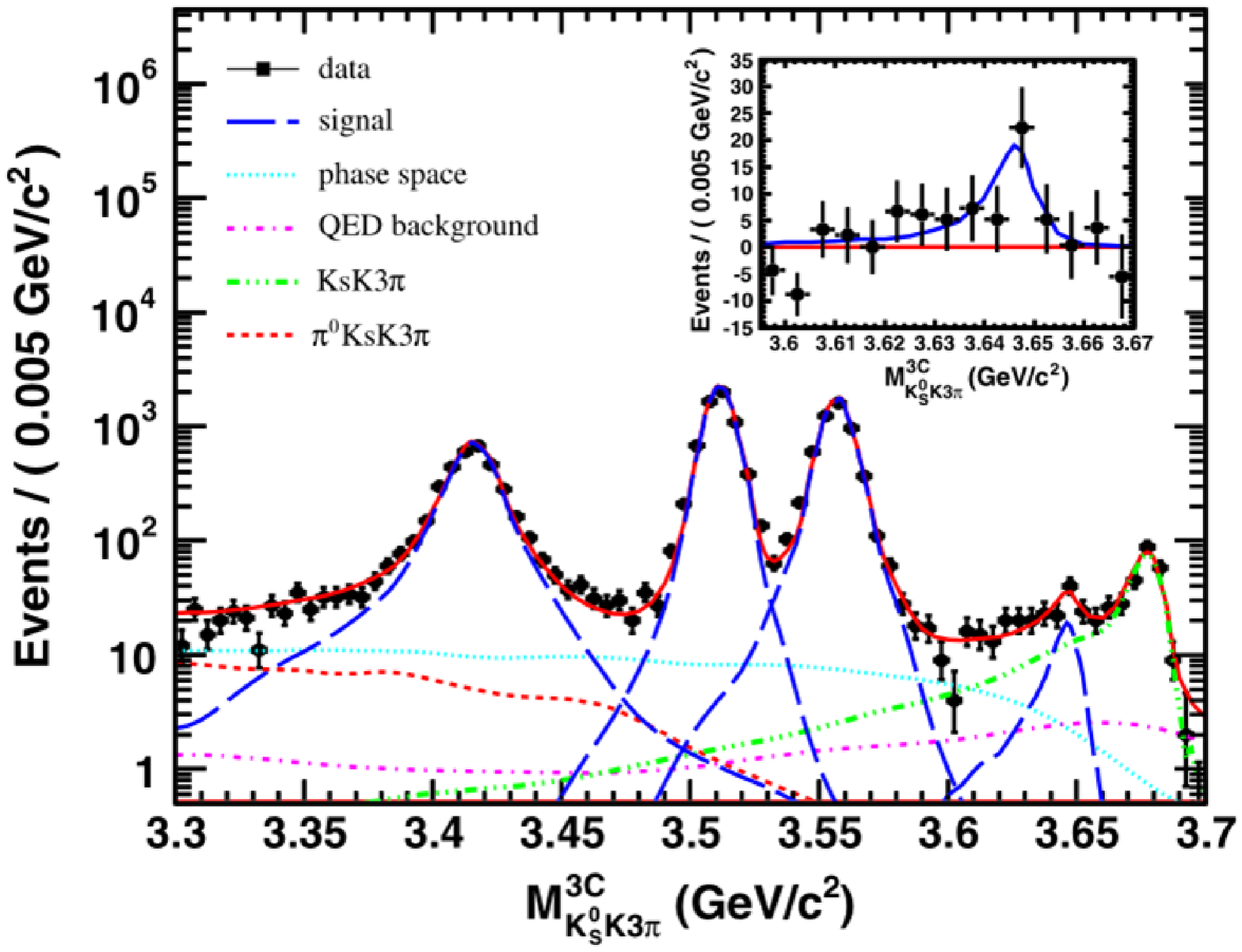}
\put(-90,100){\bf(c)}
\includegraphics[height=4cm,width=5.3cm]{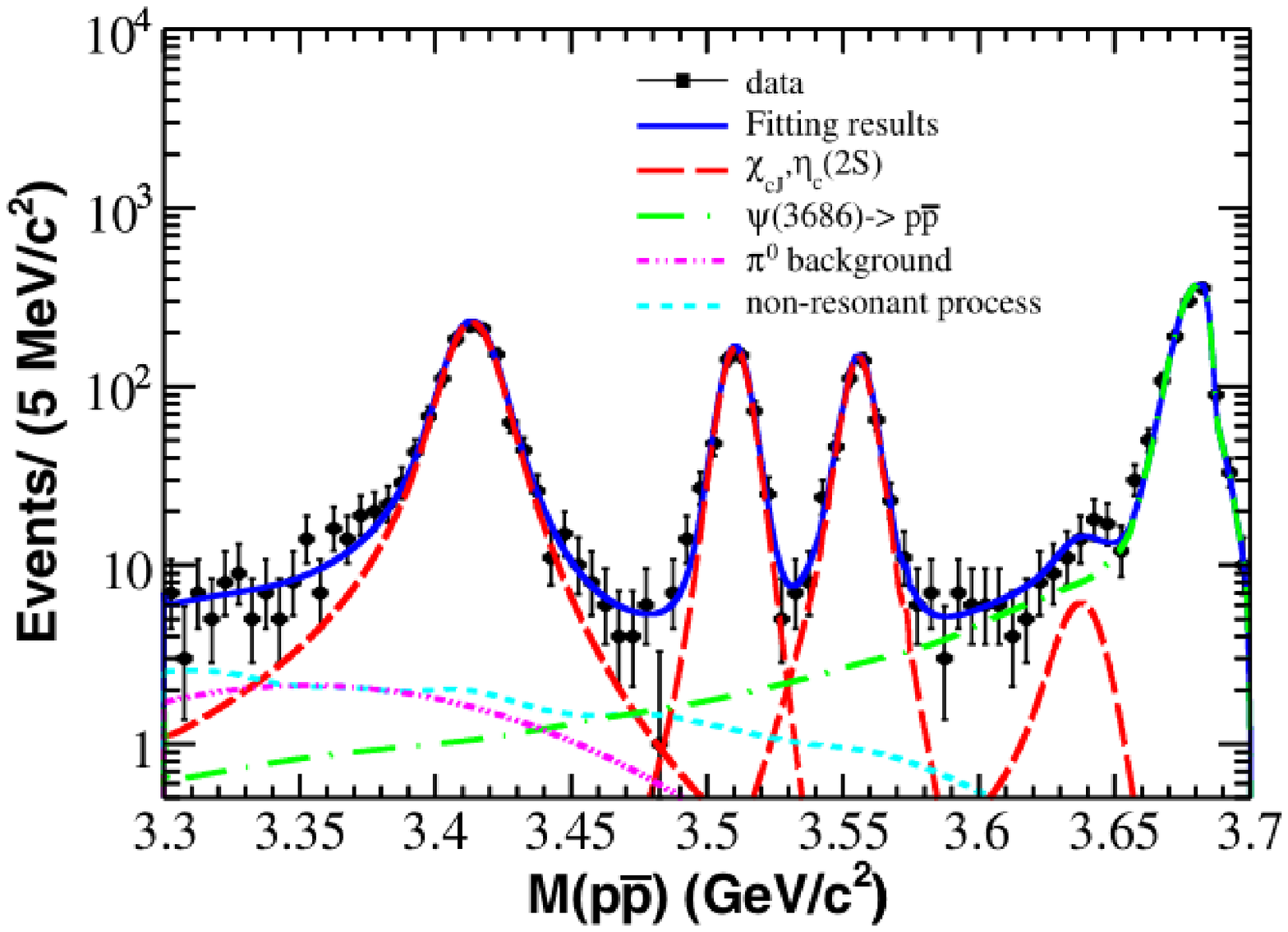}
\put(-90,100){\bf(d)}
\caption{First observation of the $M1$ transition $\psi(2S)\to\gamma\eta_c(2S)$~\cite{discoverEtac2SBES3}: the invariant-mass spectra for $K^0_SK^\pm\pi^\mp$ (a) and $K^+K^-\pi^0$ (b), as well as the simultaneous fit; Evidence of $\eta_c(2S)\to K^0_SK^{\pm}\pi^{\mp}\pi^+\pi^-$~\cite{discoverEtac2SBES32}: the invariant-mass spectra for $K^0_SK^{\pm}\pi^{\mp}\pi^+\pi^-$ (c);  Search for $\eta_c(2S)\to p\bar{p}$ via $\psi(2S)\to\gamma p\bar{p}$~\cite{etac2S2PPbar}: the invariant-mass spectra for $p\bar{p}$ (d), and the curves are total fit and each component. }
\label{etac2S}
\end{figure*}

Compared with $E1$ transitions, the rates for $M1$ transitions between charmonium states are much lower.
With the first round $\psi(2S)$ events, the $M1$ transition between the radial excited charmonium $S$-wave spin-triplet and the spin-singlet states: $\psi(2S)\to\gamma\eta_c(2S)$ was observed for the first time with
$\eta_c(2S)\to K^0_SK^\pm\pi^\mp$ and $K^+K^-\pi^0$ modes. The final $K^0_SK^\pm\pi^\mp$ and $K^+K^-\pi^0$ mass spectra and the fit results are shown in Fig.~\ref{etac2S} (a) and (b), respectively. The fit yields
for the $\eta_c(2S)$ signal events are $81\pm14$ for the $K^0_SK^\pm\pi^\mp$ mode and $46\pm11$ for the $K^+K^-\pi^0$ mode; the overall statistical significance of the signal is larger than 10$\sigma$~\cite{discoverEtac2SBES3}.

The mass of the $\eta_c(2S)$ is measured to be $(3637.6\pm2.9\pm1.6)$ MeV/$c^2$, the width $(16.9\pm6.4\pm4.8)$ MeV, in good agreement with the PDG values~\cite{pdg}, and the product BF ${\cal{B}}(\psi(2S)\to\gamma\eta_c(2S))\times{\cal{B}}(\eta_c(2S)\to K\bar{K}\pi)$ $=(1.30\pm0.20\pm0.30)\times10^{-5}$. Combining this result with a BABAR measurement of ${\cal{B}}(\eta_c(2S)\to K\bar{K}\pi)$~\cite{etac2S2KKPIBelle}, the $M1$ transition rate is determined to be ${\cal{B}}(\psi(2S)\to\gamma\eta_c(2S))$ $=(6.8\pm1.1\pm4.5)\times10^{-4}$. This agrees with theoretical calculations~\cite{preetac2S1,preetac2S2,preetac2S3} and naive estimates based on the $J/\psi(1S)\to\gamma\eta_c(1S)$ transition~\cite{searchEtac2SCLEO}.

\subsection{Evidence for $\eta_c(2S)$ in $\psi(2S)\to\gamma K^0_SK^{\pm}\pi^{\mp}\pi^+\pi^-$}
Using the same sample as mentioned above, search for the $M$1 radiative transition $\psi(2S)\to\gamma\eta_c(2S)$ by reconstructing the exclusive $\eta_c(2S)\to K^0_SK^{\pm}\pi^{\mp}\pi^+\pi^-$ decay is performed~\cite{discoverEtac2SBES32}. The final $\eta_c(2S)\to K^0_SK^{\pm}\pi^{\mp}\pi^+\pi^-$ mass spectra and the fit results are shown in Fig.~\ref{etac2S} (c).
The result for the yield of $\eta_c(2S)$ events is $57\pm17$ with a significance of 4.2$\sigma$. The measured mass of the $\eta_c(2S)$ is $(3646.9\pm1.6\pm3.6)$ MeV/$c^2$, and the width is
$(9.9\pm4.8\pm2.9)$ MeV. Comparing with BESIII previous measurement~\cite{discoverEtac2SBES3}, the width is consistent with each other within one standard deviation and the mass is about two standard deviations. The BF is measured to be ${\cal{B}}(\psi(2S)\to\gamma\eta_c(2S))$ $\times {\cal{B}}(\eta_c(2S)\to K^0_S K^{\pm}\pi^{\mp}\pi^+\pi^-)$ $=(7.03\pm2.10\pm0.70)\times10^{-6}$.
This measurement complements the previous BESIII measurement of $\psi(2S)\to\gamma\eta_c(2S)$ with
$\eta_c(2S)\to K^0_SK^\pm\pi^\mp$ and $K^+K^-\pi^0$~\cite{discoverEtac2SBES3}.

To compare with the BABAR results~\cite{etac2S12},
\begin{eqnarray}
\frac{{\cal{B}}(\eta_c(2S)\to K^+K^-\pi^+\pi^-\pi^0)}{{\cal{B}}(\eta_c(2S)\to K^0_SK^{\pm}\pi^{\mp})}=2.2\pm0.5\pm0.5, \label{eq3-2}
\end{eqnarray}
we take the value of $(4.31\pm0.75)\times10^{-6}$ as measured by BESIII for ${\cal{B}}(\psi(2S)\to\gamma\eta_c(2S))\times{\cal{B}}(\eta_c(2S)\to K^0_SK^{\pm}\pi^{\mp})$~\cite{discoverEtac2SBES3},
and assuming that
\begin{eqnarray}
\frac{{\cal{B}}(\eta_c(2S)\to K^+K^-\pi^+\pi^-\pi^0)}{{\cal{B}}(\eta_c(2S)\to K^0_SK^{\pm}\pi^{\mp}\pi^+\pi^-)}=1.52, \label{eq3-3}
\end{eqnarray}
where the value 1.52 is calculated in $\chi_{cJ}(1P)$ decays, which has the same isospin, we obtain
\begin{eqnarray}
\begin{split}
&\frac{{\cal{B}}(\eta_c(2S)\to K^+K^-\pi^+\pi^-\pi^0)}{{\cal{B}}(\eta_c(2S)\to K^0_SK^{\pm}\pi^{\mp})}\\
&=1.52\times\frac{{\cal{B}}(\eta_c(2S)\to K^0_SK^{\pm}\pi^{\mp}\pi^+\pi^-)}{{\cal{B}}(\eta_c(2S)\to K^0_SK^{\pm}\pi^{\mp})}\\
&=2.48\pm0.56\pm0.33. \label{eq3-4}
\end{split}
\end{eqnarray}
These two results are consistent with each other after considering the statistical and systematic uncertainties.

With weighted least squares method~\cite{weighted}, and combining statistical and systematic errors in quadrature, the averaged mass for $\eta_c(2S)$ at BESIII is calculated to be $(3641.45\pm2.53)$ MeV/$c^2$, and corresponding value of hyperfine splitting between $^1S_0$ and $^3S_1$ states is $\Delta{M}_{\rm hf}(2S)=(44.69\pm2.53)$ MeV/$c^2$ which agrees well with the theoretical prediction~\cite{preetac2SHPL}.

\subsection{Search for $\eta_c(2S)$ in $\psi(2S)\to\gamma p\bar{p}$}
In massless QCD models, the processes $\eta_c(1S)$/$\eta_c(2S)$/$\chi_{c0}(1P)$$\to p\bar{p}$ are forbidden by the helicity selection rule~\cite{HSL}. However, the experimental observations of the decay
$\eta_c(1S)$/$\chi_{c0}(1P)\to p\bar{p}$~\cite{pdg}, indicate substantial contributions due to finite masses. These observations have stimulated many theoretical efforts~\cite{HSL1,HSL2,HSL3}. In Ref.~\cite{HSL4}, it is pointed out that the BF of $\eta_c(2S)\to p\bar{p}$ with respect to that of $\eta_c(1P)\to p\bar{p}$ may serve as a criterion to validate the helicity
conservation theorem, and an anomalous decay in $\eta_c(2S)$ might imply the existence of a glueball.

The decays $\eta_c(2S)\to p\bar{p}$ are searched for via $\psi(2S)\to\gamma p\bar{p}$, based on 448 million $\psi(2S)$ data sample. The invariant mass $p\bar{p}$ spectrum are shown in Fig.~\ref{etac2S} (d). 
The upper limits of the product BFs are determined to be ${\cal{B}}(\psi(2S)\to\gamma\eta_c(2S))$
$\times{\cal{B}}(\eta_c(2S)\to p\bar{p})<1.4\times10^{-6}$ at the 90\% C.L.~\cite{etac2S2PPbar}.

\subsection{Search for $\eta_c(2S)$ decays into vector meson pairs}
The decay modes $\eta_c(2S)\to VV$, are supposed to be highly suppressed by the helicity selection rule~\cite{HSR}.
But in Ref.~\cite{peretac2S2VV}, a high production rate of $\eta_c(2S)\to VV$ is predicted, taking into consideration significant contributions from intermediate charmed
meson loops, which provides a mechanism to evade helicity selection rule~\cite{zhaoq2,zhaoq3}. The measurement of ${\cal{B}}(\eta_c(2S)\to VV)$ may help in understanding
the role played by charmed meson loops in $\eta_c(1S)\to VV$~\cite{peretac2S2VV}.

\begin{figure*}[hbtp]
\centering
\epsfig{width=0.3\textwidth,clip=true,file=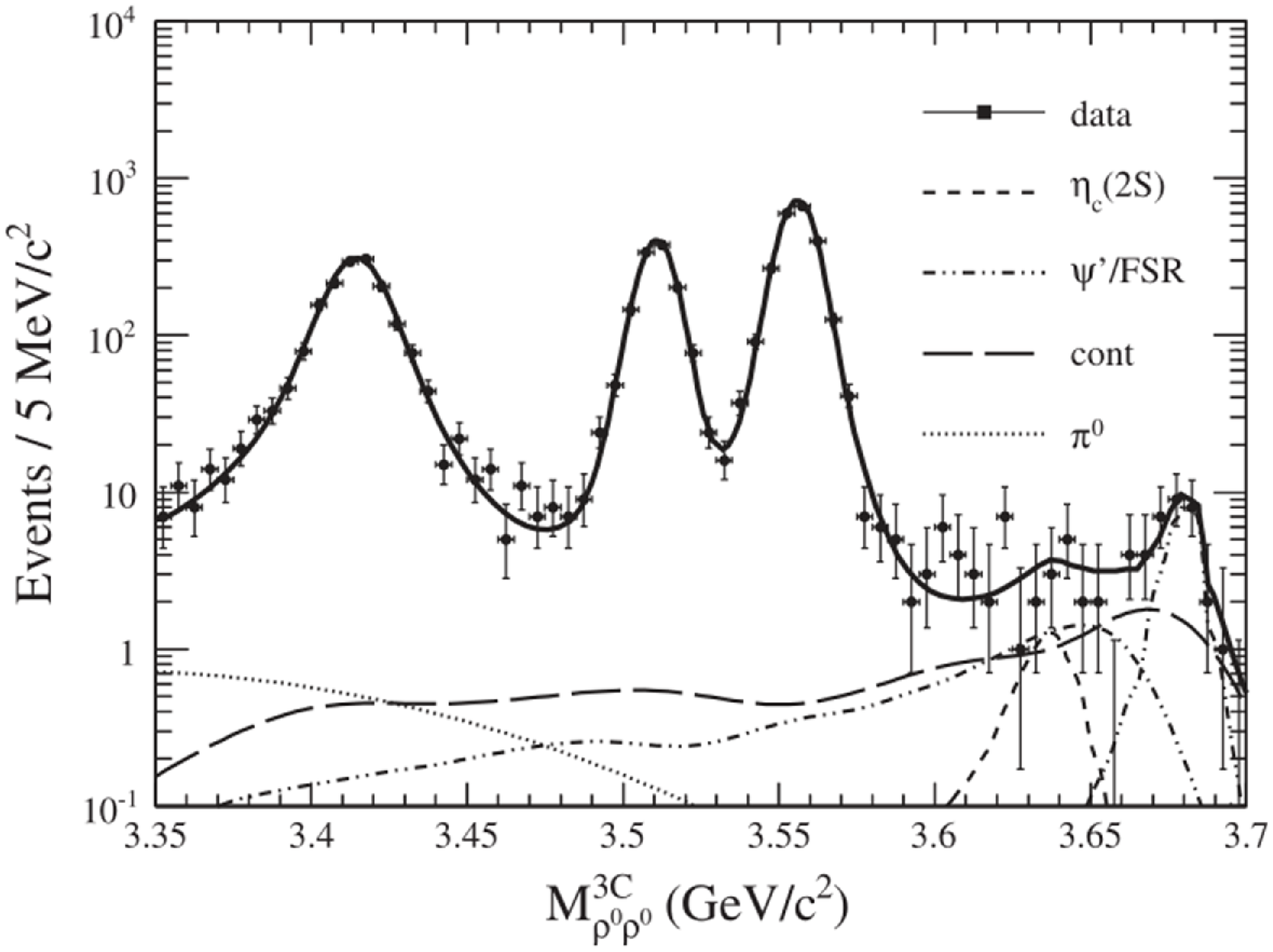}
\put(-90,100){\bf(a)}
\epsfig{width=0.3\textwidth,clip=true,file=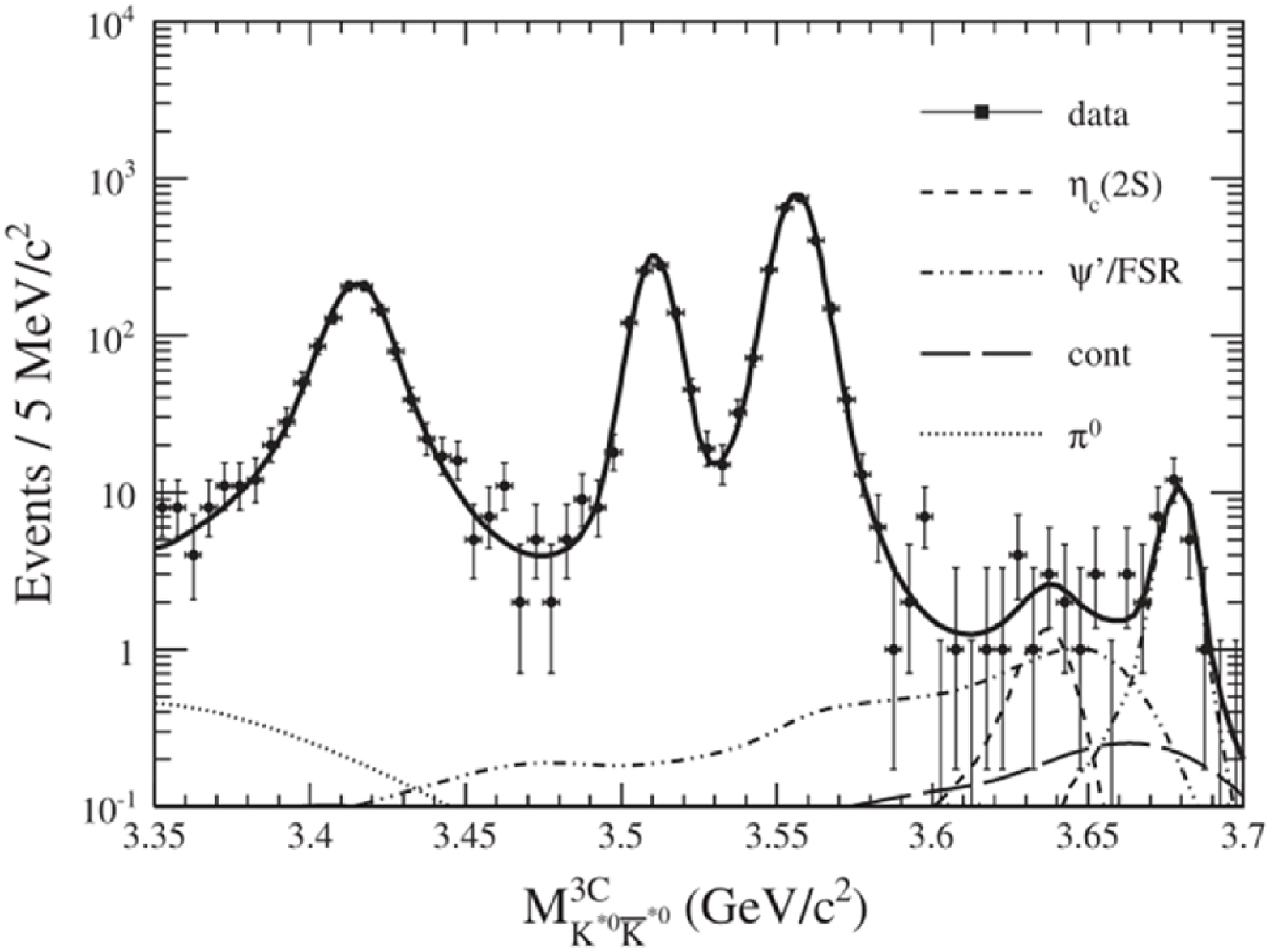}
\put(-90,100){\bf(b)}
\epsfig{width=0.3\textwidth,clip=true,file=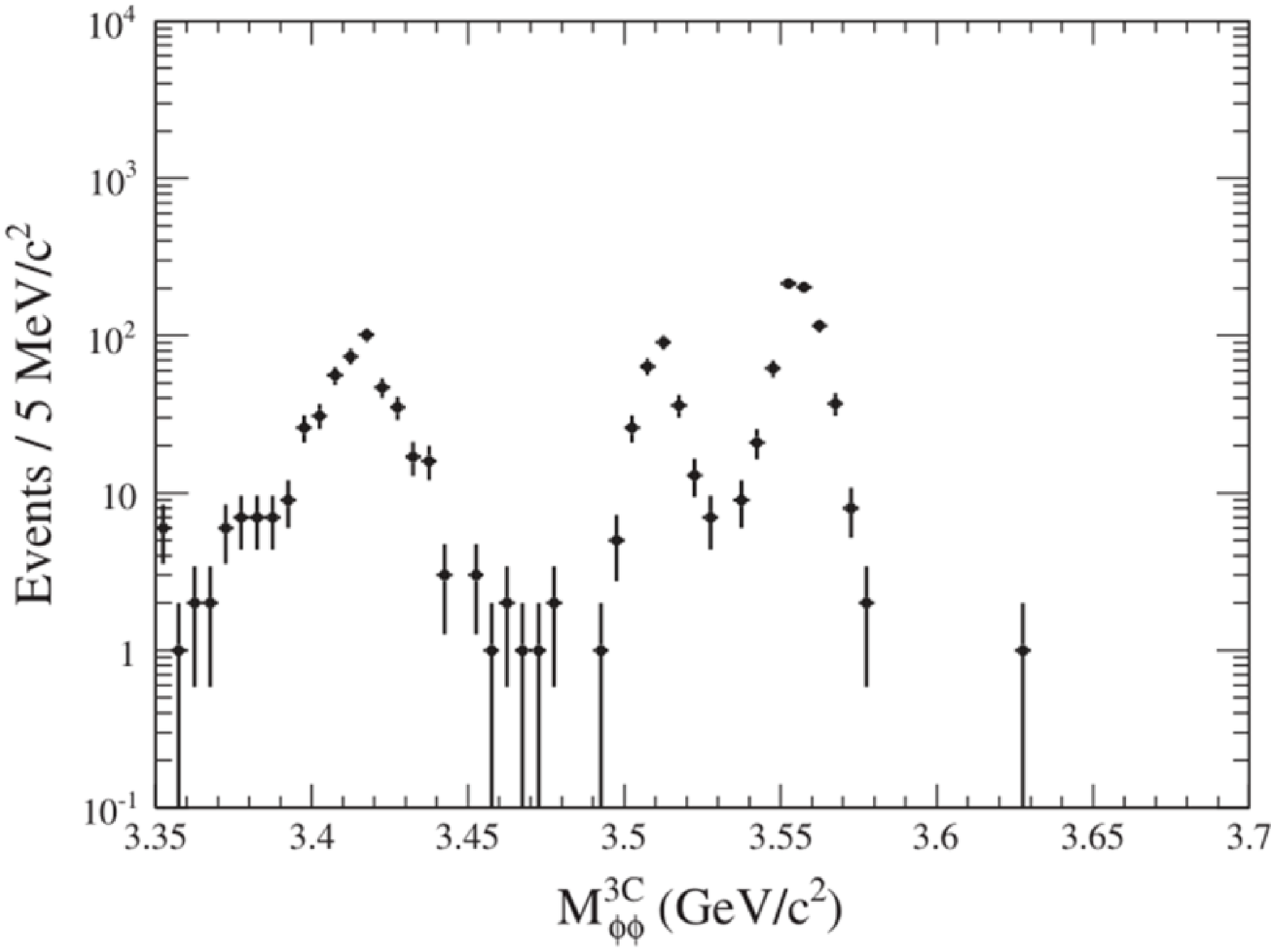}
\put(-90,100){\bf(c)}
\caption{Search for $\eta_c(2S)$ decays into vector meson pairs via $\psi(2S)\to\gamma\eta_c(2S)$~\cite{discoverEtac2SBES33}. The fit to the invariant-mass spectra for $\rho^0\rho^0$ (a), $K^{*0}\bar{K}^{*0}$ (b) and $\phi\phi$ (c). Dots with error bars are data,
The curves in (a) and (b) are total fit and each component. No fit is performed for (c) due to low statistics.}
\label{etac2S3}
\end{figure*}

The processes $\eta_c(2S)\to\rho^0\rho^0, K^{*0}\bar{K}^{*0}$ and $\phi\phi$ are searched for using a sample of $1.06\times10^8$ $\psi(2S)$ events~\cite{discoverEtac2SBES33}. The final $\rho^0\rho^0, K^{*0}\bar{K}^{*0}$ and $\phi\phi$ mass spectra and the fit results are shown in Fig.~\ref{etac2S3}. No signal are observed in any of the three decay modes. The upper limits on the decay BFs are determined to be ${\cal{B}}(\eta_c(2S)\to\rho^0\rho^0)<3.1\times10^{-3}$, ${\cal{B}}(\eta_c(2S)\to K^{*0}\bar{K}^{*0})<5.4\times10^{-3}$, and ${\cal{B}}(\eta_c(2S)\to \phi\phi)<2.0\times10^{-3}$ at 90\% C.L.. The upper limits are lower than the existing theoretical predictions~\cite{peretac2S2VV} where the predicted
BFs are ${\cal{B}}(\eta_c(2S)\to\rho^0\rho^0)=(6.4\sim28.9)\times10^{-3}$, ${\cal{B}}(\eta_c(2S)\to K^{*0}\bar{K}^{*0})=(7.9\sim35.8)\times10^{-3}$ and ${\cal{B}}(\eta_c(2S)\to \phi\phi)=(2.1\sim9.8)\times10^{-3}$, although the difference between the upper limit determined here and the existing theoretical prediction~\cite{peretac2S2VV} is very small for $\eta_c(2S)\to\phi\phi$.

\section{$h_c(1P)$ physics}
\label{sec:hcdecays}
Since its discovery at CLEO experiment in 2005~\cite{hc1,hc2}, few measurements of $h_c(1P)$ exclusive decay are available. Its dominate decay is the radiative transition $h_c(1P)\to\gamma\eta_c(1S)$~\cite{hc2getacFM,hc2getacCLEO}, while the sum of the other known $h_c(1P)$ decay BFs is less than 3\%~\cite{pdg}.

\subsection{$h_c(1P)$ resonance parameters}
The properties of the $h_c(1P)$ are theoretically investigated in Refs.~\cite{prehc1,prehc2,prehc2review,prehc3}. For example, Y. P. Kuang
considered the effect of $S$-$D$ mixing and predicted the ${\cal{B}}(\psi(2S)\to\pi^0 h_c(1P))$ to be $=(0.4-1.3)\times10^{-3}$,  he also calculated the other BFs, e.g., ${\cal{B}}(h_c(1P)\to\gamma\eta_c(1S))$ = 88\% and $\Gamma(h_c(1P))$ $=(0.51\pm0.01)$ MeV via PQCD and
${\cal{B}}(h_c(1P)\to\gamma\eta_c(1S))$ = 41\% and $\Gamma(h_c(1P))$ $=(1.1\pm0.09)$ MeV via nonrelativistic QCD (NRQCD).
Godfrey and Rosner gave a different theoretical prediction, ${\cal{B}}(h_c(1P)\to\gamma\eta_c(1S))$ = 38\%~\cite{prehc4}. A recent unquenched LQCD analysis predicted the width $\Gamma(h_c(1P))$ $=(0.601\pm0.055)$ MeV~\cite{prehc5}.

The precise measurement to $h_c(1P)$ resonant parameters is important because the comparison of its mass with the $^3P$ states ($\chi_{cJ}(1P)$) provides an essential information about the spin dependence of the $c\bar{c}$ interaction, which can be obtained by precisely measuring the $^1P$ hyperfine mass splitting $\Delta{M}_{\rm hf}(1P)\equiv$ $<M(1^3 P)>-M(1^1 P_1)$, where $<M(1^3 P_J)>$
$=[M(\chi_{c0}) + 3M(\chi_{c1}) + 5M(\chi_{c2})]/9$ $=3525.30\pm0.04$ MeV/$c^2$~\cite{prehc6} is the spin-weighted
centroid of the $^3 P_J$ mass and $M(1^1 P_1)$ is the mass of the singlet state $h_c(1P)$. A non-zero hyperfine splitting
may give indication of non-vanishing spin-spin interactions in charmonium potential models~\cite{prehc7}.

According to QCD potential models, the $c\bar{c}$ interaction in a charmonium meson can be described
with a potential that includes a Lorentz scalar confinement term and a vector Coulombic term arising from one-gluon
exchange between the quark and the antiquark. The scalar confining potential makes no contribution to the hyperfine
interaction and the Coulombic vector potential produces hyperfine splitting only for S states. This leads to the
prediction of the hyperfine or triplet-singlet splitting in the $P$ states of $\Delta{M}_{\rm hf}(1P)\equiv$ $<M(1^3 P)>-M(1^1 P_1)\simeq0$~\cite{prehc7}.

The decay mode of $\psi(2S)\to\pi^0 h_c(1P)$ is studied using three methods at BESIII:
(1) just tagging the $\pi^0$, called inclusive method~\cite{hc2getacBES3}; (2) tagging the $\pi^0$ and $E1$ gamma in the $E1$ transition decay $h_c(1P)\to\gamma\eta_c(1S)$, called $E1$ tagging method~\cite{hc2getacBES3};
(3) reconstruct $\eta_c(1S)$ with 16 exclusive hadronic decays, called exclusive method~\cite{etac1S216}.

\begin{figure}[hbtp]
\centering
\includegraphics[height=8cm,width=7.3cm]{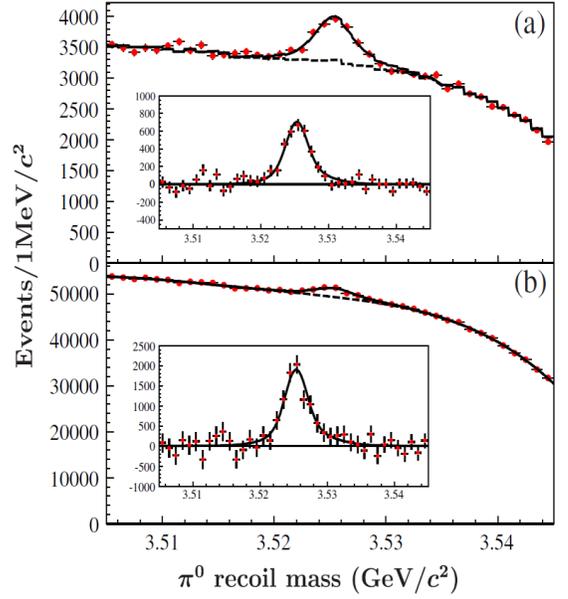}
\caption{Measurements of $h_c(1P)$ in $\psi(2S)$ decays. (a) The $\pi^0$ recoil-mass spectrum and fit for the $E1$-tagged analysis of $\psi(2S)\to\pi^0 h_c(1P)$, $h_c(1P)\to\gamma\eta_c(1S)$.
(b) The $\pi^0$ recoil-mass spectrum and fit for the inclusive analysis of $\psi(2S)\to\pi^0 h_c(1P)$. Fits are shown as solid lines,
background as dashed lines. The insets show the background-subtracted spectra~\cite{hc2getacBES3}. }
\label{hcIncl}
\end{figure}

The $\pi^0$ recoil-mass spectrum obtained with inclusive method and $E1$ tagging method, 
are shown in Fig.~\ref{hcIncl}. The corresponding event numbers of $\eta_c(1S)$ signal observed are $10353\pm1097$ and $3679\pm319$, respectively.

The absolute BFs ${\cal{B}}(\psi(2S)\to\pi^0 h_c(1P))$ $=(8.4\pm1.3\pm1.0)\times 10^{-4}$ and ${\cal{B}}(h_c(1P)\to\gamma\eta_c(1S))$ $=(54.3\pm6.7\pm5.2)$\% are measured for the first time.
A statistics-limited determination of the previously unmeasured $h_c(1P)$ width leads to an upper limit $\Gamma(h_c(1P))$
$<1.44$ MeV (90\% C.L.). The measured $M(h_c(1P))$ $=3525.40\pm0.13\pm0.18$ MeV/$c^2$ and ${\cal{B}}(\psi(2S)\to\pi^0 h_c(1P))$ $\times{\cal{B}}(h_c(1P)\to\gamma\eta_c(1S))$ $=(4.58\pm0.40\pm0.50)\times10^{-4}$ are consistent with previous results~\cite{hc2getacCLEO}.
The measured ${\cal{B}}(h_c(1P)\to\gamma\eta_c(1S))$ is close to the prediction in Ref.~\cite{prehc4} (38\%) and the NRQCD prediction of Ref.~\cite{prehc3}. The measured ${\cal{B}}(\psi(2S)\to\pi^0 h_c(1P))$ is consistent with the prediction of Ref.~\cite{prehc3}, and the total width $\Gamma(h_c)$ is consistent with the predictions of Ref.~\cite{prehc3,prehc5}.

\begin{figure}[hbtp]
\centering
\includegraphics[height=5cm,width=7.3cm]{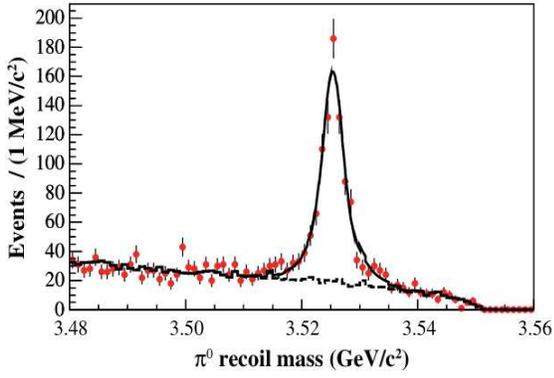}
\caption{Study of $\psi(2S)\to\pi^0 h_c(1P)$, $h_c(1P)\to\gamma\eta_c(1S)$ via $\eta_c(1S)$ exclusive decays. The $\pi^0$ recoil mass spectrum in $\psi(2S)\to\pi^0 h_c(1P)$, $h_c(1P)\to\gamma\eta_c(1S)$, $\eta_c(1S)\to X_i$ summed over the 16 final states $X_i$.
The dots with error bars represent the $\pi^0$ recoil mass spectrum in data. The solid line shows the total fit function
and the dashed line is the background component of the fit~\cite{etac1S216}. }
\label{hcExcl}
\end{figure}

The $\pi^0$ recoil-mass spectrum summed over the 16 final states, obtained with exclusive method are shown in Fig.~\ref{hcExcl}.
The measured mass and width, $M(h_c(1P))$ $=(3525.31\pm0.11\pm0.14)$ MeV/$c^2$ and $\Gamma(h_c(1P))$ $=(0.70\pm0.28\pm0.22)$ MeV, are consistent with previous measurements by $\eta_c(1S)$ inclusive decay.

With weighted least squares method~\cite{weighted}, and combining statistical and systematic errors in quadrature, the averaged mass for $h_c(1P)$ at BESIII is $(3525.35\pm0.14)$ MeV/$c^2$, and the mass splitting with $P$ wave iso-spin triplet is $\Delta{M}_{\rm hf}(1P)$ $=(-0.05\pm 0.15)$ MeV/$c^2$ which is consistent with zero with one standard deviation and not violates the assumption that there is only short-distance contribution in potential model~\cite{prehc7}.

\subsection{$h_c(1P)$ decays}

\subsubsection{Observation of $h_c(1P)\to\gamma\eta'$ and evidence for $h_c(1P)\to\gamma\eta$}
Since the $h_c(1P)$ has negative $C$ parity, it very likely decays into a photon plus a pseudoscalar meson, such as $\eta$ and $\eta^\prime$~\cite{hc2getaetap}. The $\eta$ and $\eta^{\prime}$ mesons are commonly understood as mixtures of the pure SU(3) flavor octet $[(u\overline{u}+d\overline{d}-2s\overline{s})/\sqrt{6}]$
and singlet $[(u\overline{u}+d\overline{d}+s\overline{s})/\sqrt{3}]$ states and a small gluonium component~\cite{etaetapmix1}~\cite{etaetapmix2}.
The flavor content of each $\eta^{(\prime)}$ mass eigenstate can be quantified with a mixing angle, the value of which becomes manifest in ratios of branching fractions for
various radiative decays involving a $\eta$ or $\eta^{\prime}$~\cite{etaetapmix3}~\cite{etaetapmix4}~\cite{etaetapmix5}.
The ratio of the BF ${\cal{B}}(h_c(1P)\to\gamma\eta)$ over ${\cal{B}}(h_c(1P)\to\gamma\eta^\prime)$ can be used to
study the $\eta-\eta^\prime$ mixing angle, and is also important to test SU(3)-flavor symmetries in QCD~\cite{hc2getaetapPre}.
As in the case of $\psi$ (including $J/\psi$ and $\psi(2S)$) decays, the process $\psi\to\gamma\eta$ and $\psi\to\gamma\eta^{\prime}$ occur primarily through radiative of the photon from the $c$ quark or $\overline{c}$
quark in the initial state. Assuming such a mechanism and the applicability of SU(3) symmetry for the decay amplitudes, the decay proceeds through the SU(3)-singlet part of the pseudoscalar. One finds~\cite{etaetapmix}
\begin{eqnarray}
\frac{\Gamma(\psi\to\gamma\eta^{\prime})}{\Gamma(\psi\to\gamma\eta)}
=(\frac{k_{\eta^{\prime}}}{k_{\eta}})^3\frac{1}{\rm tan^2\theta},\label{eq4-0}
\end{eqnarray}

\begin{figure*}[hbtp]
\centering
\epsfig{width=0.8\textwidth,clip=true,file=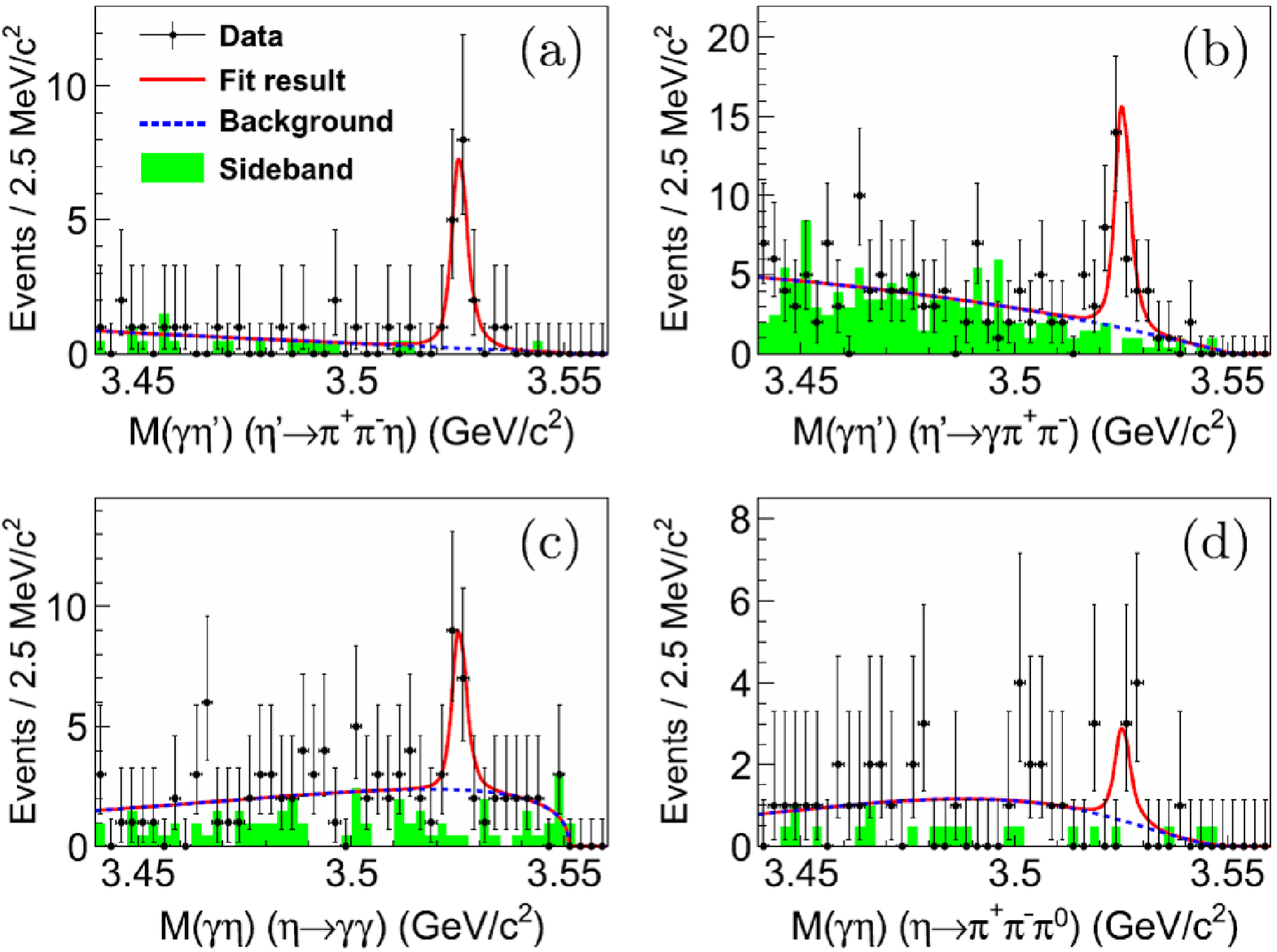}\\
\caption{Observation of $h_c(1P)\to\gamma\eta'$ and evidence for $h_c(1P)\to\gamma\eta$~\cite{hc2getaetap}. Results of the simultaneous fits to the two invariant mass distributions of (top) $M(\gamma\eta^\prime)$ and (below)
$M(\gamma\eta)$ for data. (a) $M(\gamma\eta^\prime)$ distribution for $h_c(1P)\to\gamma\eta^\prime$ $(\eta^\prime\to\pi^+\pi^-\eta)$.
(b) $M(\gamma\eta^\prime)$ distribution for $h_c(1P)\to\gamma\eta^\prime$ $(\eta^\prime\to\gamma\pi^+\pi^-)$. (c) $M(\gamma\eta)$
distribution for $h_c(1P)\to\gamma\eta$ $(\eta\to\gamma\gamma)$. (d) $M(\gamma\eta)$ distribution for $h_c(1P)\to\gamma\eta$
$(\eta\to\pi^+\pi^-\pi^0)$. The red solid curves are the fit results, the
blue dashed curves are the background distributions, and the
green hatched histograms are events from the $\eta(\eta^\prime)$ sidebands.}
\label{hc2GamEta}
\end{figure*}

The radiative decay $h_c(1P)\to\gamma\eta^\prime$ is observed with a statistical significance of 8.4$\sigma$ for the first time, and the evidence for the process $h_c(1P)\to\gamma\eta$
with a significance of 4.0$\sigma$~\cite{hc2getaetap} at BESIII. Here, $\eta^\prime$ is reconstructed with $\eta\pi^+\pi^-$ and $\gamma\pi^+\pi^-$ modes, while $\eta$ is reconstructed with $\gamma\gamma$ and $\pi^+\pi^-\pi^0$ modes.
Figure~\ref{hc2GamEta} shows the distributions of $M(\gamma\eta^\prime)$ and $M(\gamma\eta)$ for the selected events.
The corresponding BFs of $h_c(1P)\to\gamma\eta^\prime$ and $h_c(1P)\to\gamma\eta$ are
measured to be $(1.52\pm0.27\pm0.29)\times10^{-3}$ and $(4.7\pm1.5\pm1.4)\times10^{-4}$, respectively, where the first errors are statistical and the second are systematic. The ratio 
is
$R_{h_c(1P)}$ $={{\cal{B}}(h_c(1P)\to\gamma\eta)}/{{\cal{B}}(h_c(1P)\to\gamma\eta^\prime)}$ $=(30.7\pm11.3\pm8.7)$\%, where
the common systematic errors 
cancel.
The $\eta-\eta^\prime$ mixing angle extracted from $R_{h_c(1P)}$ is determined to be $27.2^\circ\pm14.9^\circ$ which is to test SU(3)-flavor symmetries in QCD~\cite{hc2getaetapPre},
following the methods used for equivalent decays of the $\psi$ mesons~\cite{calhc1,calhc2,calhc3}.

\subsubsection{Measurement of $h_{c}(1P)\to \rm light~ hadrons$}\label{sec:hc2LH}
\begin{figure}[hbtp]
\centering
\epsfig{width=0.4\textwidth,clip=true,file=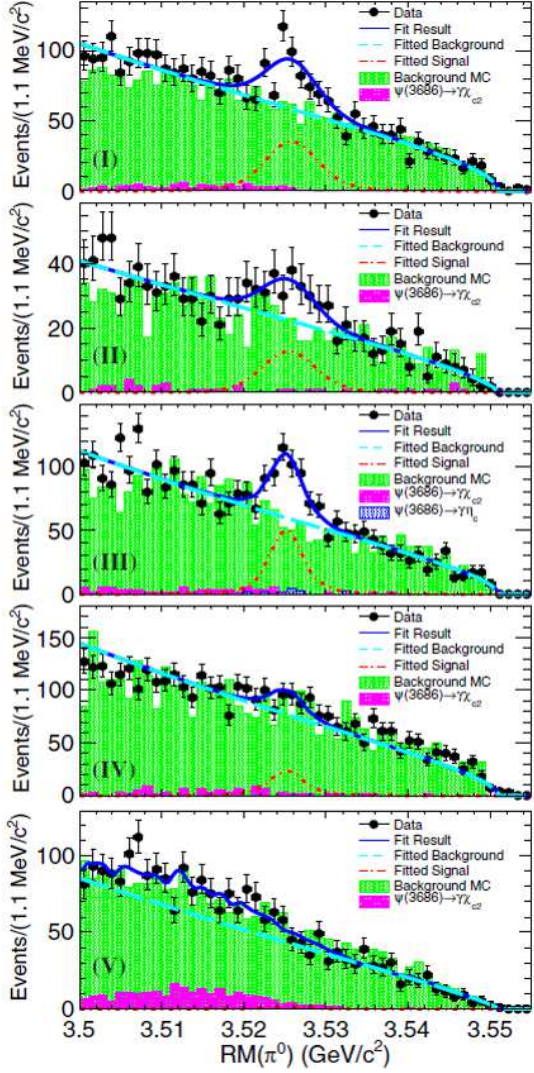}\\
\caption{Observation of $h_c(1P)\to\rm hadrons$. Recoiling mass spectra of the lowest energy $\pi^0$, in the decay chains $\psi(2S)\to\pi^0 h_c(1P)$ with $h_c(1P)\to p\bar{p}\pi^+\pi^-$ (I),
 $\pi^+\pi^-\pi^0$ (II), $2(\pi^+\pi^-)\pi^0$ (III), $3(\pi^+\pi^-)\pi^0$ (IV), and $K^+ K^-\pi^+\pi^-$ (V).  In
each spectrum, the dots with error bars represent data, the pink shaded histogram is the background process $\psi(2S)\to\gamma\chi_{c2}(1P)$,
the blue filled histogram is the background process $\psi(2S)\to\pi^0 h_c(1P)$, $h_c(1P)\to\gamma\eta_c(1S)$, the green filled histogram is the
background from inclusive MC, the cyan dashed curve is the fitted background, the red dash-dotted curve is the fitted signal, and the blue curve is the fitted result~\cite{wljhc}.}
\label{hcWlj}
\end{figure}

\begin{figure*}[hbtp]
\centering
\epsfig{width=0.9\textwidth,clip=true,file=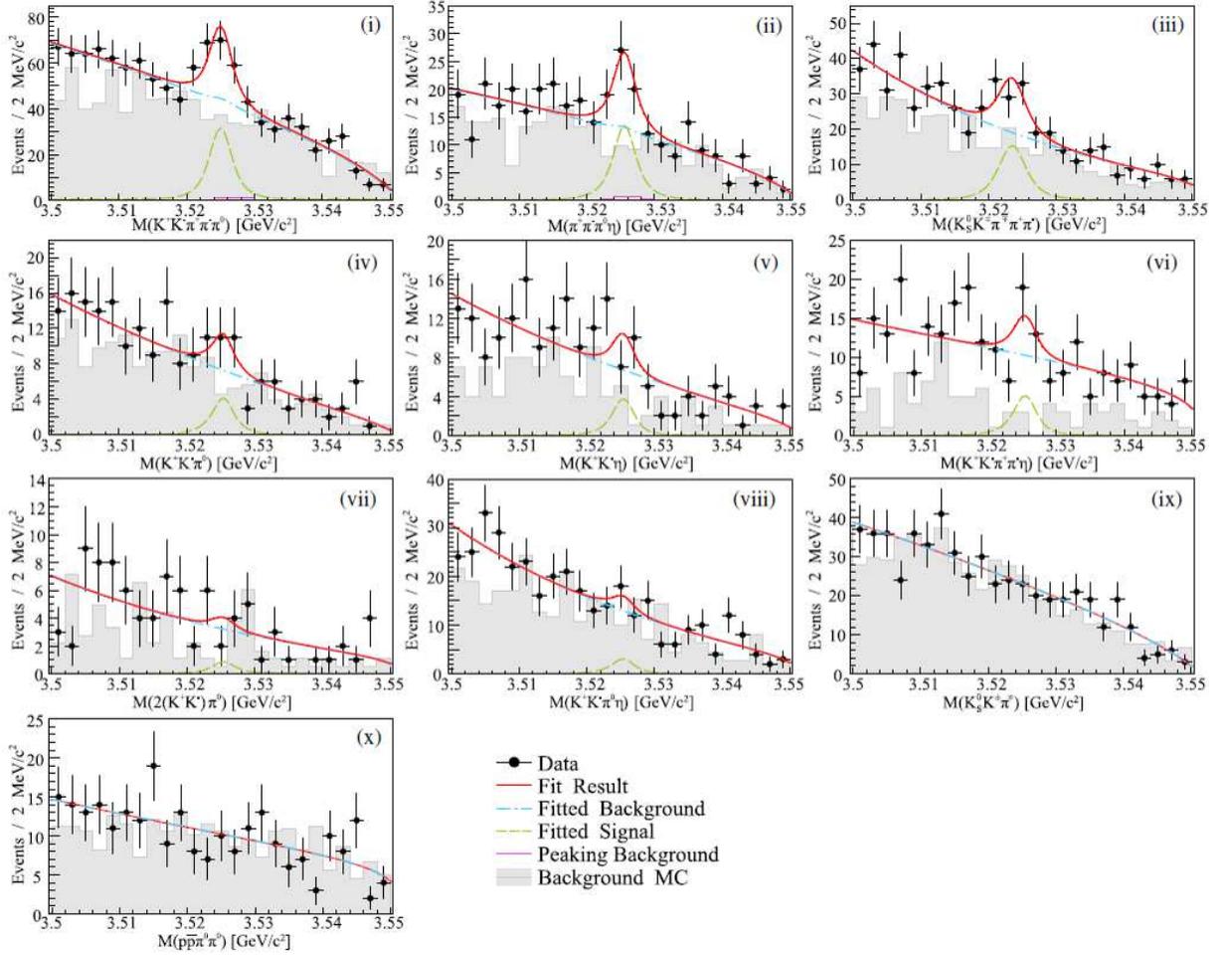}\\
\caption{Observation of $h_c(1P)\to K^+K^-\pi^+\pi^-\pi^0$. Fits to the invariant mass distributions for the $h_c$ decay modes (i)-(x). Data are shown as black points, the total fit result is
shown in red, the background contribution is denoted by the blue dashed-dotted line (including peaking background contributions for
channel (i) and (ii) as shown in magenta), the signal contribution is illustrated by the green dashed line. The background level obtained
from inclusive MC is shown by the gray shaded histogram~\cite{mikehc}.}
\label{hcMike}
\end{figure*}

For the $h_c(1P)$ hadronic decays, the predictions on the ratios of hadronic with of the $h_c(1P)$ to that of $\eta_c(1S)$, based on the PQCD and NRQCD are very different~\cite{prehc2}. Assuming that
\begin{eqnarray}
\frac{\Gamma(h_c(1P)\to h)}{\Gamma(\eta_c(1S)\to h)}
\thickapprox\frac{\Gamma(h_c(1P)\to 3g)}{\Gamma(\eta_c(1S)\to 2g)},\label{eq4-1}
\end{eqnarray}
and taking into account of
\begin{eqnarray}
\frac{\Gamma(\eta_c(1S)\to 2g)}{\Gamma(J/\psi(1S)\to 3g)}
=\frac{27}{5(\pi^2-9)\alpha_s}(\frac{M^2_{J/\psi(1S)}}{M^2_{\eta_c(1S)}}),\label{eq4-2}
\end{eqnarray}
${\Gamma(h_c(1P)\to h)}/{\Gamma(\eta_c(1S)\to h)}$ is calculated to be $0.010\pm0.001$ in PQCD, while $0.083\pm0.018$ in NRQCD,
as is the corresponding ratio involving decays of $J/\psi(1S)$ mesons $(\Gamma^{\rm had.}_{h_c(1P)}/\Gamma^{\rm had.}_{J/\psi(1S)})$.
New studies of $h_c(1P)$ hadronic decays will enable these ratios to be measured, and comparisons to be made with the
theoretical predictions. Fifteen $h_c(1P)$ exclusive hadronic decays as listed in Table.~\ref{tab::hchadrons}, are searched for via the process $\psi(2S)\to\pi^0 h_c(1P)$~\cite{wljhc,mikehc} at BESIII. Four of them, $h_c(1P)\to p\bar{p}\pi^+\pi^-$, $h_c(1P)\to\pi^+\pi^-\pi^0$, $2(\pi^+\pi^-)\pi^0$ and $K^+ K^-\pi^+\pi^-\pi^0$ are observed for the first time with significance of 7.4$\sigma$, 4.6$\sigma$, 9.1$\sigma$ and 6.0$\sigma$.
Evidences for the decays $h_c(1P)\to\pi^+\pi^-\pi^0\eta$ and $h_c(1P)\to K^0_S K^\pm \pi^\mp\pi^+\pi^-$ is found with a significance of 3.6$\sigma$ and 3.8$\sigma$, respectively. The invariant mass for these final states and the fit status are shown in Fig.~\ref{hcWlj},~\ref{hcMike}. The corresponding BRs (and upper limits) are summarized in Table.~\ref{tab::hchadrons}.

\begin{table*}[!hbtp]
\begin{center}
\caption{Review of the measured BF of $h_c(1P)\to X$ at BESIII.
Here, $X$ denotes hadronic final states, $\epsilon$ denotes the selection efficiency, $N_{h_c(1P)}$ denotes the $h_c(1P)$ signal yield. ${\cal{B}}$ denote the BF
${\cal{B}}(h_c(1P)\to\rm hadrons)$. S.S. is the significance of the signal peak, including systematic uncertainties.
The last column gives the status before the BESIII experiment.}
\begin{tabular}{llcccccc}
\hline
Decay mode   & ~~~~~$X$                  & Yield         & $\epsilon$(\%)& ${\cal{B}}(10^{-3})$ &S.S  & Ref. & Comment \\
\hline
(I)&$h_c(1P)\to p\bar{p}\pi^+\pi^-$    & $230\pm25$    & 20.9          & $2.89\pm0.32\pm0.55$ &7.4$\sigma$  &  ~\cite{wljhc} & first measurement\\
(II)&$h_c(1P)\to \pi^+\pi^-\pi^0$       & $101\pm25$    & 16.8          & $1.60\pm0.40\pm0.32$ &4.6$\sigma$  &  ~\cite{wljhc} & $<2.2$\\
(III)&$h_c(1P)\to 2(\pi^+\pi^-)\pi^0$    & $254\pm32$    & 9.1           & $7.44\pm0.94\pm1.52$ &9.1$\sigma$  &  ~\cite{wljhc} & $22^{+8}_{-7}$\\
(IV)&$h_c(1P)\to 3(\pi^+\pi^-)\pi^0$    & $73\pm34$     & 4.2           & $4.65\pm2.17\pm1.08$ &2.1$\sigma$  &  ~\cite{wljhc} & $<29$\\
&                               & $<136$        &               & $<8.7$                 &             &                & \\
(V)&$h_c(1P)\to K^+K^-\pi^+\pi^-$      & $<40$         & 18.1          & $<0.6$                 & ...         &  ~\cite{wljhc} &first measurement\\\hline
(i)&$h_c(1P)\to K^+K^-\pi^+\pi^-\pi^0$ &$80\pm15$      & 6.5           & $3.3\pm0.6\pm0.6$    & 6.0$\sigma$ &  ~\cite{mikehc}&first measurement\\
(ii)&$h_c(1P)\to \pi^+\pi^-\pi^0\eta$   &$35\pm9$       & 3.3           & $7.2\pm1.8\pm1.3$    & 3.6$\sigma$ &  ~\cite{mikehc}&\multirow{2}{*}{first measurement}\\
    &                           & $<50.0$         &               & $<18$                &             &                &\\
(iii)&$h_c(1P)\to K^0_S K^{\pm}\pi^{\mp}\pi^+\pi^-$&$41\pm13$& 5.5       & $2.8\pm0.9\pm0.5$    & 3.8$\sigma$ &  ~\cite{mikehc}&\multirow{2}{*}{first measurement}\\
    &                           & $<65.3$       &               & $<4.7$               &             &                &\\
(iv)&$h_c(1P)\to K^+K^-\pi^0$           & $<20.1$       & 9.8           & $<0.6$               & ...         &  ~\cite{mikehc}&first measurement\\
(v)&$h_c(1P)\to K^+K^-\eta$            & $<18.5$       & 14.3          & $<0.9$               & ...         &  ~\cite{mikehc}&first measurement\\
(vi)&$h_c(1P)\to K^+K^-\pi^+\pi^-\eta$  & $<24.1$       & 6.9           & $<2.5$               & ...         &  ~\cite{mikehc}&first measurement\\
(vii)&$h_c(1P)\to 2(K^+K^-)\pi^0$        & $<11.7$       & 6.7           & $<0.3$               & ...         &  ~\cite{mikehc}&first measurement\\
(viii)&$h_c(1P)\to K^+K^-\pi^0\eta$       & $<20.2$       & 6.3           & $<2.2$               & ...         &  ~\cite{mikehc}&first measurement\\
(ix)&$h_c(1P)\to K^0_S K^{\pm}\pi^{\mp}$& $<17.4$       & 14.4          & $<0.6$               & ...         &  ~\cite{mikehc}&first measurement\\
(x)&$h_c(1P)\to p\bar{p}\pi^0\pi^0$    & $<11.8$       & 8.7           & $<0.5$               & ...         &  ~\cite{mikehc}&first measurement\\
\hline
\end{tabular}
\label{tab::hchadrons}
\end{center}
\end{table*}

Table~\ref{tab::BRhcdecays} shows the comparisons of the measured ratios of the hadronic decay widths $\Gamma^{\rm had.}_{h_c(1P)}/\Gamma^{\rm had.}_{\eta_c(1S)}$ and $\Gamma^{\rm had.}_{h_c(1P)}/\Gamma^{\rm had.}_{J/\psi(1S)}$
and the theoretical predictions. The experimental results tend to favor the lower predictions, which come from pQCD. However, in Ref.~\cite{prehc3}, the theoretical prediction of
${\cal{B}}(h_c(1P)\to\gamma\eta_c(1S))=(41\pm3)$\% based on NRQCD is favored by the experimental measurement $(51\pm6)$\%~\cite{pdg}, compared with the prediction of $(88\pm2)$\% from pQCD.
 We note that the experimental measurements are still limited by low statistics and the predictions of the theoretical models can be modified through considerations such
 as normalization scale or relativistic corrections~\cite{corrPrehc1}~\cite{corrPrehc2}. Future experimental measurements of higher precision, and improved theoretical
  calculations will help to resolve this inconsistency.

\begin{table}[!hbtp]
\begin{center}
\caption{The ratios of hadronic decay width of $h_c(1P)$ to $\eta_c(1S)$ ($\Gamma^{\rm had.}_{h_c(1P)}/\Gamma^{\rm had.}_{\eta_c(1S)}$) and $h_c(1P)$ to $J/\psi(1S)$ ($\Gamma^{\rm had.}_{h_c(1P)}/\Gamma^{\rm had.}_{J/\psi(1S)}$).
The theoretical predictions of the total hadronic decay ratios are based on pQCD and NRQCD~\cite{prehc3}, which are expected to be correct also for exclusive decay modes. The experimental measurements of the
ratios of the partial decay widths for $p\overline{p}\pi^+\pi^-$, $K^+K^-\pi^+\pi^-$, and $n(\pi^+\pi^-)\pi^0$ ($n = 0,1,2$) modes are calculated based on the measured branching fractions in Ref.~\cite{wljhc}~\cite{mikehc} and the PDG~\cite{pdg}.
}
\begin{tabular}{lcc}
\hline
   &    Model/mode   & Ratio\\
\hline
\multirow{5}*{$\Gamma^{\rm had.}_{h_c(1P)}/\Gamma^{\rm had.}_{\eta_c(1S)}$}         & PQCD    & $0.010\pm0.001$      \\
                                                                            & NRQCD   & $0.083\pm0.018$  \\
                                                                            & $p\overline{p}\pi^+\pi^-$   & $0.012\pm0.008$  \\
                                                                            & $K^+K^-\pi^+\pi^-$          & $< 0.002$  \\
                                                                            & $K^+K^-\pi^+\pi^-\pi^0$    & $0.002\pm0.001$      \\
                                                                            &$K^0_S K^{\pm}\pi^{\mp}$ & $<0.001$\\
                                                                            & $K^0_S K^\pm\pi^\mp\pi^\pm\pi^\mp$    & $<0.002$      \\
                                                                            & $K^+K^-\pi^0$    & $<0.001$      \\
                                                                            & $K^+K^-\eta$    & $<0.001$      \\
                                                                            & &\\
\multirow{8}*{$\Gamma^{\rm had.}_{h_c(1P)}/\Gamma^{\rm had.}_{J/\psi(1S)}$}         & PQCD    & $0.68\pm0.07$      \\
                                                                            & NRQCD   & $8.03\pm1.31$      \\
                                                                            & $p\overline{p}\pi^+\pi^-$   & $3.63\pm2.25$      \\
                                                                            & $\pi^+\pi^-\pi^0$   & $0.57\pm0.38$      \\
                                                                            & 2($\pi^+\pi^-)\pi^0$   & $1.43\pm0.90$      \\
                                                                            & 3($\pi^+\pi^-)\pi^0$   & $<2.26$      \\
                                                                            & $K^+K^-\pi^+\pi^-$     & $<0.66$      \\
                                                                             & $K^+K^-\pi^0$    & $<2.35$      \\
                                                                             &$K^0_S K^{\pm}\pi^{\mp}$ & $<0.81$\\
                                                                            & $K^+K^-\pi^+\pi^-\pi^0$     & $2.07\pm1.40$       \\
                                                                            & $K^+K^-\pi^+\pi^-\eta$ & $<4.0$ \\
\hline
\end{tabular}
\label{tab::BRhcdecays}
\end{center}
\end{table}

\subsubsection{Search for $h_c(1P)\to\pi^+\pi^- J/\psi$}
Hadronic transitions between the heavy quarkonium ($Q\bar{Q}$) states are particularly interesting for  testing the interplay between PQCD and NPQCD~\cite{preQQYQ}. A common approach for calculating these transitions is the QCD multipole expansion~\cite{preQQYQ2} for gluon emission.
The calculation depends on experimental inputs and works well for transitions of heavy $Q\bar{Q}$ states below open flavor threshold~\cite{preQQYQ3}. However to date, the only well-measured hadronic transitions in the charmonium sector are those for the $\psi(2S)$.

For charmonium states below the $D\bar{D}$ threshold, the hadronic transitions of the spin-singlet P-wave state $h_c(1P)$ are one of the best places to test the spin-spin interaction between heavy quarks~\cite{prehcYQ}, but they remain the least accessible experimentally because the $h_c$ cannot be produced resonantly in $e^+ e^-$ annihilation or from
electric-dipole radiative transitions of the $\psi(2S)$.

The $h_c(1P)$ is expected to decay to lower-mass charmonium state through hadronic transitions, but this has not been observed yet. In the framework of QCDME, the BF of $h_c(1P)\to\pi\pi J/\psi(1S)$  (including charged and neutral modes)is predicted to be 2\%~\cite{prehcYQ2}, while it is predicted to be 0.05\% when neglecting the
nonlocality in time~\cite{prehcYQ3}.  An experimental measurement is desirable to distinguish between these calculations.

A search for the hadronic transition $h_c(1P)\to\pi^+\pi^- J/\psi(1S)$ is carried out via $\psi(2S)\to\pi^0 h_c(1P)$, $h_c(1P)\to\pi^+\pi^- J/\psi(1S)$ at BESIII~\cite{hc2ppijpsi}. No signal is observed. The upper limit of the product of
BFs ${\cal{B}}(\psi(2S)\to\pi^0 h_c(1P))$ ${\cal{B}}(h_c(1P)\to\pi^+\pi^- J/\psi(1S))$ at the 90\% C.L. is determined to be $2.0\times10^{-6}$. Using the PDG value for the BF
of $\psi(2S)\to\pi^0 h_c(1P)$ of $(8.6\pm1.3)\times10^{-4}$~\cite{pdg}, the upper limit on ${\cal{B}}(h_c(1P)\to\pi^+\pi^- J/\psi(1S))$ is determined to be $2.4\times10^{-3}$, which is the most  stringent upper limit to date. Neglecting the small phase space difference between the charged and neutral $\pi\pi$ modes and assuming isospin symmetry, upper limit for ${\cal{B}}(h_c(1P)\to\pi\pi J/\psi(1S))$ at 90\% C.L. is determined to be 3.6$\times10^{-3}$ (including charged and neutral modes). It is noted that the measured BF is smaller than the prediction in Ref.~\cite{prehcYQ2} by one order in magnitude, but does not contradict that in Ref.~\cite{prehcYQ3}.

\section{Summary and prospectives}
\label{sec:summary}
With the capability of adjusting the $\EE$ c.m. energy to the peaks of resonances, combined with the clean experimental environments due to near-threshold operatopn,
BESIII is uniquely able to perform a broad range of critical measurements of charmonium physics, as discussed above in the context of the studies of $\eta_c(1S)$, $\eta_c(2S)$ and $h_c(1P)$ states.

Despite the impressive progress, many $\eta_c(1S)$, $\eta_c(2S)$ and $h_c(1P)$ decays are still to be observed and explored.

At present, the BESIII detector has collected the world's largest $J/\psi(1S)$ data sample of $N_{J/\psi(1S)}= (10087\pm44)\times10^{6}$~\cite{jpsitot}.
In addition, BESIII has collected a sample of $2.55\times10^9$ $\psi(2S)$ data sample in this year. Thus, there will be about $3.0\times10^9$ $\psi(2S)$ data sample in total. BESIII also has a great plan for collecting the $XYZ$ data sample~\cite{newpsip}.

With these unique advantage of an unprecedented high-statistics data samples, the BESIII experiment could further study the $\eta_c(1S)$, $\eta_c(2S)$ and $h_c(1P)$ state in the following aspects.

\indent 1. Give more precise measurements of the masses and widths of the $\eta_c(1S)$ and $h_c(1P)$ via $E1$ transition $h_c(1P)\to\gamma\eta_c(1S)$ because of the negligible interference and $\eta_c(2S)$ via the $M1$ transition $\psi(2S)\to\gamma\eta_c(2S)$ with statistical uncertainty reduced significantly, and better understand the line shapes associated with their production, to precisely test the predicted hyperfine mass splitting based on the potential models~\cite{prehc7}, and recent lattice computations~\cite{LQCD1,LQCD2,LQCD3}, as well as quark-model predictions~\cite{qmpre}.

\indent 2. Give more precise measurement for their known hadronic decay modes, and investigate more decay modes of the $\eta_c(1S)$, $\eta_c(2S)$ and $h_c(1P)$ states to test the predictions on the ratios of hadronic width of the $h_c(1P)$ to that of $\eta_c(1S)$,  and the ratios of hadronic decay width of $h_c(1P)$ to that of $J/\psi(1S)$, based on pQCD and NRQCD~\cite{prehc3}. For instance, $h_c(1P)\to p\bar{p}\pi^+\pi^-\pi^0$, $p\bar{p}\eta$ and $p\bar{p}\pi^0$, $\eta_c(2S)\to\pi^+\pi^-\eta$, and so on.

\indent 3. Search for the Di-pion transition decays $\eta_c(2S)\to\eta_c(1S)\pi^+\pi^-$ and $h_c(1P)\to\pi^+\pi^- J/\psi(1S)$.
In particular, the Di-pion transition amplitude for $\eta_c(2S)\to\eta_c(1S)\pi^+\pi^-$ is expected~\cite{PrePiPietac1S} to have the same approximately linear dependence on the squared invariant mass of the di-pion system as the $\psi(2S)\to\ J/\psi(1S)\pi^+\pi^-$~\cite{Psi2STPiPiJpsi}. PHSP integration of the squared amplitude, evaluated for the peak masses $M_{\eta_c(1S)}$ and $M_{\eta_c(2S)}$ of the $\eta_c(1S)$ and $\eta_c(2S)$, respectively, yields $\Gamma(\eta_c(2S)\to\eta_c(1S)\pi^+\pi^-)$/$\Gamma(\psi(2S)\to J/\psi(1S)\pi^+\pi^-)\approx$ 2.9. This will leads to the BF prediction ${\cal{B}}(\eta_c(2S))\to\eta_c(1S)\pi^+\pi^-$ $=(2.2^{+1.6}_{-0.6})$\%. This decay may be further suppressed due to the contribution of the chromomagnetic interaction to the decay amplitude~\cite{PrePiPietac1S2}.
However, it is not observed yet in experiment yet~\cite{PiPietac1S}~\cite{ExpPiPietac1S}, and only the upper limit of its branching fraction is determined and to be ${\cal{B}}(\eta_c(2S)\to\eta_c(1S)\pi^+\pi^-) <$ 25\%~\cite{pdg}.

\indent 4. Study the electromagnetic (EM) Dalitz decay $h_c(1P)\to e^+e^-\eta_c(1S)$ and $J/\psi(1S)/\psi(2S)\to\EE\eta_c(1S)$, which have access to the EM transition form factors (TFFs) of these charmonium states.
The $q^2$ dependence of charmonium TFFs can provide additional information of the interactions between the charmonium states and the electromagnetic field, where $q^2$ is the square of
the invariant mass of the $\EE$ pair, and serve as a sensitive probe to their internal structures. Furthermore, the $q^2$-dependent TFF can possibly distinguish the transition mechanisms based on the $c\bar{c}$ scenario and other solutions which alter the simple quark model picture.

\indent 5. Study the $\eta_c(1S)$ state via the $M1$ transition to improve the precision of ${\cal{B}}(\psi(2S)\to\gamma\eta_c(1S))$
, and the $\eta_c(2S)$ state with ${\cal{B}}(\psi(2S)\to\gamma\eta_c(2S))$ which could be used to extract the absolute BFs for some specific $\eta_c(2S)$ decays.  There are other radiative transitions such as $\eta_c(2S)\to\gamma J/\psi(1S)$, $\eta_c(2S)\to\gamma h_c(1P)$, $\chi_{c2}(1P)\to\gamma h_c(1P)$, and $h_c(1P)\to\gamma\chi_{c0,1}(1P)$ that are challenges for BESIII even with $10^9$ $\psi(2S)$ data sample because of low decay rates or difficulty in detecting the soft photon, but these rates can be calculated in the potential model~\cite{presoftgamdecay} and experimental searches are therefore important.

\indent 6.
Study the radiative decays of the charmonium states for a better understanding of charmonium decay dynamics, such as $\eta_c(1S)\to\gamma V$ with the predicted BR in the level of $10^{-6}\sim10^{-7}$ in the framework of NRQCD~\cite{preetac2gamV}.
And the improved measurement of $\eta_c(1S)\to\gamma\gamma$ will shed light on the effects of higher order QCD corrections as well as
provide validation of the decoupling of the hard and soft contributions in the NRQCD framework due to its simplicity~\cite{preetac2gg}.
With the larger data samples, all of these measurements will be improved.

\indent 7. Study the two-body baryonic decays of the $\eta_c(1S)$, $\eta_c(2S)$ and $h_c(1P)$, which can provide information on color-siglet and color-octet contribution.
The $\eta_c(2S)$ and $h_c(1P)$ decaying into $p\bar{p}$ have been searched for with $1\times10^8$ $\psi(2S)$ decays sample at BESIII and no obvious signal has been observed. The upper limit of the BFs are set to be $3.0\times10^{-3}$ and $1.5\times10^{-4}$, respectively. With the larger $\psi(2S)$ data sample of $3\times10^9$, we could find the evidence of them with the assumption that the efficiency is about 40\% and there is no background.

\indent 8. Dalitz plot analysis of $\eta_c(1S)\to K^+K^-\eta$, $\eta_c(1S)\to K^+K^-\pi^0$, $\eta_c(1S)\to K^+K^-\eta^{\prime}$, $\eta_c(1S)\to\pi^+\pi^-\eta^{\prime}$ and $\eta_c(1S)\to\pi^+\pi^-\eta$, in order to investigate the light meson spectroscopy in which scalar mesons remain a puzzle with the reason that they have complex structure, and there are too many states to be accommodated within the quark model without difficulty~\cite{etacLMS}. Decays of the $\eta_c(1S)$ provide a window on light meson states. A series of related work has been carried out with two photons processes by BABAR~\cite{twophotons}. The world's largest $J/\psi(1S)$ data sample of $N_{J/\psi(1S)}= (10087\pm44)\times10^{6}$ at BESIII provides the good chance for this topic via the M1 transition $J/\psi(1S)\to\gamma\eta_c(1S)$.

\begin{acknowledgements}
\label{sec:acknowledgement}
This work is supported in part by the National Natural Science Foundation of China (11605042),
basic research plan for key scientific research projects of higher education institutions in Henan Province (21A140012),
cultivation Fund for National scientific research projects of Henan Normal University (2021PL07),
excellent Youth Foundation of Henan Province (212300410010),
the youth talent support program of Henan Province (ZYQR201912178),
the National Natural Science Foundation of China (11875122,11735014),
and the Program for Innovative Research Team in University of Henan Province (19IRTSTHN018). This paper is published in a preprint~\cite{preprint}.

\end{acknowledgements}

\end{document}